\documentclass[twoside,11pt, preprint]{article}

\usepackage{blindtext}

\usepackage{jmlr2e}

\newcommand{\bX}{\mathbf{X}}
\newcommand{\bR}{\mathbf{R}}
\let\hat\widehat

\usepackage{lastpage}

\firstpageno{1}

\begin{document}

\title{Markov Missing Graph: A Graphical Approach for Missing Data Imputation}

\author{\name Yanjiao Yang \email yjyang00@uw.edu \\
       \addr Department of Statistics\\
       University of Washington\\
       Seattle, WA 98195-4322, USA
       \AND
       \name Yen-Chi Chen \email yenchic@uw.edu \\
       \addr Department of Statistics\\
       University of Washington\\
       Seattle, WA 98195-4322, USA}


\maketitle

\begin{abstract}
We introduce the Markov missing graph (MMG), a novel framework that  imputes missing data based on undirected graphs. MMG leverages conditional independence relationships to locally decompose the imputation model. To establish the identification, we introduce the Principle of Available Information (PAI), which guides the use of all relevant observed data. We then propose a flexible statistical learning paradigm, MMG Imputation Risk Minimization under PAI, that frames the imputation task as an empirical risk minimization problem. This framework is adaptable to various modeling choices. We develop theories of MMG, including the connection between MMG and Little's complete-case missing value assumption, recovery under missing completely at random,  efficiency theory, and graph-related properties. 
We show the validity of our method with simulation studies and illustrate its application with a real-world Alzheimer's data set.
\end{abstract}

\begin{keywords}
  imputation, missing data, missing not at random,  undirected graph, semi-parametric efficiency 
\end{keywords}

\section{Introduction}
Missing data are common problems in applied research, arising in domains as diverse as the social sciences \citep{molenberghs2014handbook}, biomedical studies \citep{bell2014handling}, and machine learning applications \citep{emmanuel2021survey}. When unaccounted for, missingness can introduce bias, inflate variance, and compromise generalization, undermining both statistical inference and predictive performance. These issues highlight the importance of principled methodologies for handling missing data in modern data analysis and machine learning pipelines.

Recent years have witnessed substantial advances in graphical approaches for addressing missing data. In particular, directed acyclic graphs (DAGs) are widely used to encode independence assumptions and establish identification results \citep{mohan2013graphical, mohan2014testability, tian2015missing, shpitser2015missing, bhattacharya2020identification, nabi2020full}.  A broad class of models can be represented by DAGs, including the permutation model \citep{robins1997MNAR} and the block-sequential missing-at-random (MAR) model \citep{zhou2010block}. \cite{mohan2021graphical} reviewed recent developments in this area and proposed testable implications for DAGs. 
Beyond DAGs, chain graphs (CGs) and acyclic directed mixed graphs (ADMGs) have also been applied to missing data problems. CGs have been used to represent missing not at random (MNAR) models such as the no self-censoring model \citep{shpitser2016consistent, sadinle2017itemwise, malinsky2022semiparametric} and the self-censoring model \citep{li2023self}. \cite{nabi2020full} built connections by showing that a missing data model of DAGs or ADMGs satisfying certain graphical conditions is a submodel of the no self-censoring model.
More recently, \cite{chen2022pattern} introduced a new concept termed ``pattern graphs'', which provides a general representation encompassing identifying restrictions such as \citet{molenberghs1998monotone}, \citet{tchetgen2018discrete}, and \cite{chen2019nonparametric}. \cite{phung2025recursiveequationsimputationmissing} proposed a novel characterization for identifying the full data law in graphical models and introduced a new imputation algorithm that can handle missingness patterns with no support. 
These graphical approaches have established identification theory and enabled principled analysis of missing data. 
However, existing graphical approaches predominantly focus on DAGs, ADMGs, or CGs, leaving undirected graphs largely unexplored for missing data problems. 

Imputation is a widely used strategy for handling missing data, where unobserved entries are replaced with plausible values drawn from a model. This remains an active research area across statistics \citep{murray2016multiple, suen2023modeling} and machine learning \citep{angelopoulos2023prediction}. Classical examples include multiple imputation by chained equations (MICE) \citep{raghunathan2001multivariate, burgette2010multiple, JSSv045i03} and the nonparametric, random forest–based missForest method \citep{stekhoven2012missforest}. More recent developments leverage neural architectures to capture nonlinear dependencies and scale computation, such as GAIN \citep{yoon2018gain}, MIWAE \citep{pmlr-v97-mattei19a}, and MIDA \citep{gondara2018mida}. Importantly, imputation is rarely an end in itself: for example, multiple imputation can be interpreted as a Monte Carlo approximation to regression adjustment.
Despite their practical success, most imputation methods are primarily motivated by algorithmic design and empirical performance, with comparatively little attention to theoretical guarantees. In particular, questions of identifiability, efficiency, and consistency remain underexplored. This gap highlights the need for frameworks that not only perform well in practice but also admit rigorous statistical analysis.



The review of prior work highlights two critical and interconnected gaps. First, the use of undirected graphs in missing data problems remains underexplored, despite their importance in statistical modeling and machine learning. Second, most existing imputation methods lack a rigorous theoretical foundation. These gaps motivate the development of a unified framework that integrates undirected graphs with imputation while providing formal guarantees such as identifiability, recovery, and efficiency.


In this paper, we propose the Markov missing graph (MMG), a novel framework for handling missing data. The MMG leverages undirected graph structures to facilitate imputation by decomposing the imputation model into submodels. To the best of our knowledge, this paper is the first to introduce an undirected graph-based framework for missing data that provides both theoretical guarantees and principled imputation strategies. 

\emph{Main Results.} Our main contributions are as follows:
\begin{itemize}
    \item We propose an imputation model based on the MMG and the principle of available information (PAI) that  nonparametrically identify the full-data distribution (Section \ref{sec::MMG}). 
    \item We introduce a statistical learning paradigm (IRM: imputation risk minimization), which formulates the estimation of imputation model as an empirical risk minimization problem. This formulation allows MMG to serve as a unified framework that accommodates flexible modeling choices (Section \ref{sec::learning}).
    \item We show that MMG and PAI reduces to the CCMV \citep{little1993pattern} under two special cases (Theorems~\ref{thm: PAI_CCMV} and \ref{thm: PAI_CCMV_monotone}) and recover the correct graphical model under MCAR (Theorem \ref{thm: recovery underMCAR}).
    \item We derive the inverse probability weighting, regression adjustment, and multiply-robust estimator
    based on MMG and PAI (Section \ref{sec::efficiency}). 
    We also study the related efficiency theory (Theorems~\ref{thm: EIF} and \ref{thm: multiple robustness}).
    \item We investigate the graphical properties for the MMG imputation model (Section \ref{sec::graph}).
    \item We use both simulation data and an analysis on the National Alzheimer's Coordinating Center data to demonstrate the effectiveness of MMG and PAI (Sections \ref{sec::simulation} and \ref{sec::nacc}).  
    
\end{itemize}

\emph{Outline.} The remainder of this paper is organized as follows. 
Section \ref{sec::MMG} presents MMG and the principle of available information (PAI), which lead to nonparametric identification. In Section \ref{sec::learning}, we develop a statistical learning framework, with formally defined MMG imputation risk and illustrative parametric examples.  Section \ref{sec::theory} establishes theoretical properties of MMG, including the connection between PAI and CCMV, recovery under MCAR, efficiency theory, and graph-related properties. Section \ref{sec::simulation} provides simulation studies that compare MMG with existing imputation methods. In Section \ref{sec::nacc} we apply MMG to real-world Alzheimer's data from the NACC study and conduct sensitivity analysis. 
Proofs
and technical details are in the Appendix. We develop an R package \texttt{mmg} for implementation (\url{https://github.com/yjyang00/mmg}).

\subsection{Notations and Graph Theory}
Let $X \in \mathbb{R}^d$ be a vector of study variables subject to missingness and $R \in \{0,1\}^d$ be a binary vector indicating the response pattern of $X$. Variable $R_j = 1$ if $X_j$ is observed and $R_j = 0$ otherwise. Let $\bar{R} = \mathbf{1}_d - R$ be the reverse (flipping 0 and 1) of pattern $R$, where $\mathbf{1}_d = (1,1,\cdots,1)^T \in \mathbb{R}^d$ is the response pattern corresponding to the complete cases. For notation convenience, let $R=r$ be shorthand for $(R_1,\dots,R_d)=(r_1,\dots,r_d)$ and $R=1$ be the case of $R_1=R_2=\cdots=R_d=1$.
For a response pattern $R=r$, we use the notation $X_r = (X_j : r_j = 1)$ to denote the observed entries. Then $X_{\bar{r}}$ will be the variables of $X$ that are missing under the pattern $R=r$.

For two response patterns $r_1, r_2$, we define $r_1>r_2$ if $r_{1,j}\geq r_{2,j}$ for all $j\in \{1,\dots,d\}$ and there exists at least one index $k$ such that $r_{1,k}>r_{2,k}$. 
For example, suppose that $X=(X_1,X_2,X_3)^T\in \mathbb{R}^3$, then $X_{101}$ = $(X_1,X_3)^T$, $X_{001}=X_3$, and $X_{\overline{001}}=(X_1,X_2)^T$. It holds that $\text{101}>\text{001}$ and $\text{101}>\text{100}$, but the pattern $\text{101}$ is not comparable with $\text{011}$ or $\text{010}$. It follows directly from the definition that if $r_1>r_2$, the set of observed variables under pattern $r_2$ is a subset of those observed under pattern $r_1$.

The observed data are independent identically distributed (IID) realizations of the vector $(R,X_{R})$. The notation \( p(\cdot) \) is used for distributions or density functions. 
To avoid abuse of notations, we use the boldface
$$
(\bX_1,\bR_1),\cdots, (\bX_n, \bR_n)
$$
to denote the variables representing our data (oracle data).
The observed data is then the IID collections
$$
(\bX_{1, \bR_1},\bR_1),\cdots, (\bX_{n, \bR_n}, \bR_n).
$$

A graph comprises vertices and edges that encode conditional independence relations among a collection of variables. In this study, we consider an undirected graph $G = (V, E)$, where $V$ is a set of vertices corresponding to study variables and $E$ is a set of undirected edges. The vertices $u$ and $v$ are adjacent in $G$ if there exists an edge $e(u, v) \in E$.  The neighbors of a vertex $v$ are defined as $N_G(v) = \{u \in V : e(u,v) \in E\}$, the set of all vertices adjacent to $v$. For a set of variables $U \subset V$, let $N_G(U)$ be the union of neighbors excluding variables in $U$, \ie, $N_G(U) = \{v \in V \setminus U : \exists\, u \in U \text{ such that } e(u, v) \in E\}$. For convenience, the operation $N_G$ also applies to a binary vector $r \in \{0,1\}^d$. Specifically, we define $N_G(r) = N_G(V_r)$, where $V_r = (j : r_j = 1)$ is the set of observed variables in $V$ under the pattern $R=r$. Also, for any $r$, we denote $\bar N_G(r) = N_G(r) \cup r$, where $N_G(r) \cup r \equiv\{j: j \in N_G(r) \mbox{ or } r_j =1\}$ is the indices of variables in either the boundary $N_G(r)$ or the response vector $r$. A connected subset of $G$ is a collection of vertices $A\subset V$ such that vertices in $A$ are connected in $G$. A connected component of $G$ is a connected subset such that it is not a subset of any other connected subset. For a response vector $r$, a connected subset (or component) of $r$ refers to the connected subset (or component) of the subgraph of $G$ with vertices restricted to $V_r$.


\section{Markov Missing Graph}  \label{sec::MMG}

Based on the above notations, we can decompose the joint probability density/mass function (PDF/PMF) of $(X,R)$ as 
$$
p(x,r) = p(x_{\bar r}|x_r, R=r) p(x_r,R=r),
$$
where $p(x_{\bar r}|x_r,R=r)$ is called the \emph{extrapolation density} that characterizes the unobserved variables' distribution given the observed variables and $p(x_r,R=r)$
is called the observed-data density since it describes the observed part of the data. This decomposition is also known as the pattern mixture model decomposition \citep{little1993pattern}.
An imputation model can be viewed as a model of the extrapolation density $p(x_{\bar r}|x_r, R=r)$. 
To see why an imputation model is a model of $p(x_{\bar r}|x_r,R=r)$, consider a simple example where $X=(X_1,X_2,X_3)^T$ and suppose our observe data is $X = (x_1,\texttt{NA}, x_3)$, which corresponds to the response pattern $R=101$. 
Then the extrapolation density will be $p(x_2|X_1=x_1,X_3=x_3, R=101)$,
which is exactly the density of the unobserved variable $X_2$ given the observed variables.

A Markov missing graph (MMG) uses an undirected graph $G = (V, E)$ that represent the information on the construction of an imputation model, where the vertex set $V=\{X_1,\dots,X_d\}$ consists of variables and the edge set $E$ specifies conditional independence.

\begin{figure}
    \centering
    \includegraphics[width=\linewidth]{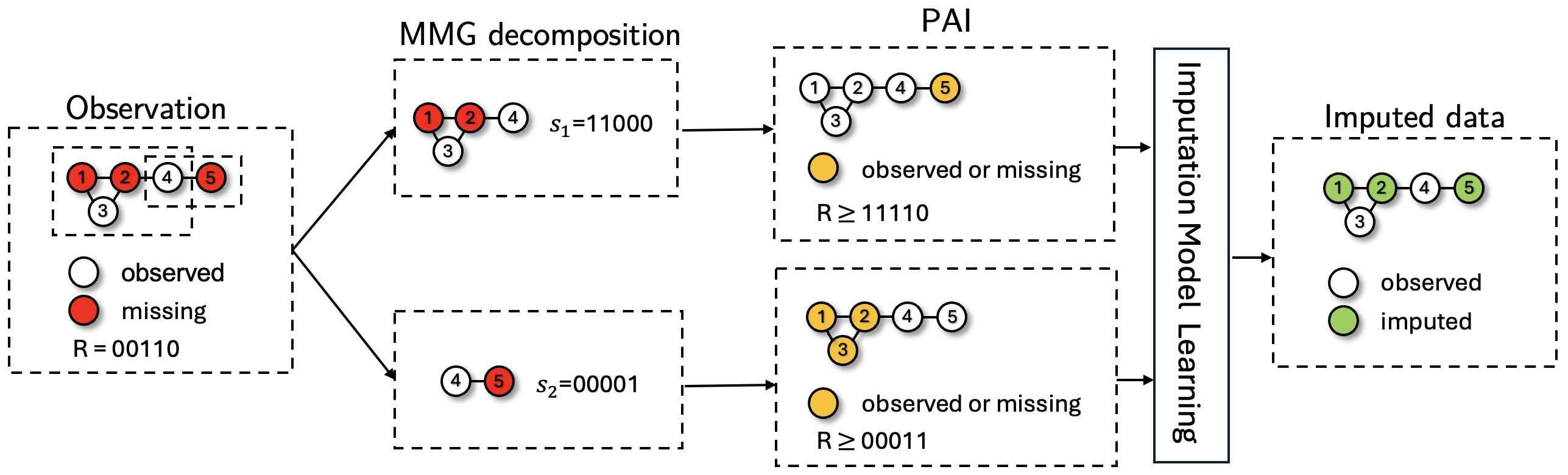}
    \caption{Overview of the proposed Markov missing graph framework for missing data imputation. In this example, the graph $G$ has five nodes (variable $X_1$--$X_5$) and edges $\{(1,2), (1,3), (2,3), (2,4), (4,5)\}$.}
    \label{fig:example1}
\end{figure}

\textbf{Markov property.} The idea of MMG is as follows. The imputation model for a missing variable $X_j$ is assumed to use only information from its neighborhood in $G$. Since $G$ is an undirected graph, this locality constraint leads to an imputation with a \textit{Markov property}. Formally, given a response pattern $r$, let $s_1, \dots, s_K$ be the binary vectors representing the connected components of $\bar{r}$ (missing variables) in the graph $G$. Namely, $\bar{r} = s_1 + s_2 + \cdots + s_K$. MMG decomposes the imputation model as
\begin{equation}
p(x_{\bar{r}} | x_r, R = r) = \prod_{k=1}^K p(x_{s_k} | x_{N_G(s_k)}, R = r).
\label{eq::MMG1}
\end{equation}
For example, consider the graph $G$ in Figure \ref{fig:example1}. Given the response pattern $r=00110$, the missing variables are $X_{\bar{r}}=X_{11001}=(X_1,X_2,X_5)^T$. The binary vectors for the connected components of $\bar{r}$ are $s_1=11000$ and $s_2=00001$. Consequently, MMG decomposes the imputation model as $p(x_1,x_2,x_5 | x_3,x_4,R=00110)=p(x_1,x_2|x_3,x_4,R=00110)p(x_5 | x_4,R=00110)$. 

\textbf{Principle of available information.} The submodel $p(x_{s_k}|x_{N_G(s_k)}, R = r)$ is not identifiable because variables $x_{s_k}$ are missing under response pattern $R = r$. To identify this model, we introduce the \textit{principle of available information} (PAI). PAI states that, without further information, one should use all available information to identify the model $p(x_{s_k} | x_{N_G(s_k)}, R = r)$. The relevant variables for this submodel are $x_{\bar{N}_G(s_k)} = x_{s_k} \cup x_{N_G(s_k)}$, where we define $\bar{N}_G(s_k) = N_G(s_k) \cup s_k$ to be the collection of $s_k$ and its neighbors. PAI states that any observation 
with variables $x_{\bar{N}_G(s_k)}$ being observed should be used to construct the imputation model. 

\subsection{Nonparametric MMG (NP-MMG)} 
Under a nonparametric model, we do not specify any parametric form for the imputation model $p(x_{s_k}|x_{N_G(s_k)}, R = r)$. The PAI leads to the  following identifying restriction
\begin{equation}
p(x_{s_k}|x_{N_G(s_k)}, R = r) = p(x_{s_k}|x_{N_G(s_k)},
 R_{\bar{N}_G(s_k)} =1),
\label{eq::MMG2}
\end{equation}
where \( R_{\bar{N}_G(s_k)} = 1 \) means that \( R_j = 1 \) for all \( j \in \bar{N}_G(s_k) \). The model $p(x_{s_k}|x_{N_G(s_k)}, R_{\bar{N}_G(s_k)} = 1)$
is clearly identifiable because all relevant variables are observed in this case. This leads to the following theorem.

\begin{theorem}
\label{thm: nonp-identify}
The MMG and PAI 
(equations~\eqref{eq::MMG1} and~\eqref{eq::MMG2}) nonparametrically identify the full-data distribution $p(x,r)$.
\end{theorem}

In the terminology of \citet{robins1997MNAR} and \citet{vansteelandt2006ignorance}, 
Theorem \ref{thm: nonp-identify} establishes what is known as nonparametric saturated models \citep{robins1997MNAR} or nonparametric identifiability \citep{vansteelandt2006ignorance}. 
Note that if we have additional information, we can further restrict to the subset of individuals within the group of \( R_{\bar{N}_G(s_k)} = 1 \). PAI states that if there is no further information, we should use all available individuals, which in this case refers to those with \( R_{\bar{N}_G(s_k)} = 1 \).

Note that MMG and PAI are clearly not the MAR assumption. Therefore, they form an identifying assumption belonging to the class of missing-not-at-random (MNAR). 

\begin{example}
\label{example1} Consider the graph in Figure~\ref{fig:example1} with response pattern $r=00110$, where variables $X_1,X_2,X_5$ are missing. MMG decomposes the imputation model as \[p(x_1,x_2,x_5 | x_3,x_4,R=00110)=p(x_1,x_2|x_3,x_4,R=00110)p(x_5 | x_4,R=00110).\] Therefore, we need to identify the submodels \( p(x_1, x_2 | x_3, x_4, R = 00110) \) and \( p(x_5 | x_4, R = 00110) \). By PAI, the first submodel can be identified using all observations where the variables $X_1,X_2,X_3,X_4$ are observed. 
Namely, PAI implies that
\[
p(x_1, x_2 | x_3, x_4, R = 00110) = p(x_1, x_2 | x_3, x_4, R_1 = R_2 = R_3 = R_4 = 1) \equiv p(x_1, x_2 | x_3, x_4, R \geq 11110).
\]
where $R\geq 11110$ denotes patterns in which at least $X_1$ through $X_4$ are observed. That is, observations with $R_i\in \{11110, 11111\}$ are included.
Similarly, for the second submodel, PAI implies that
\[
p(x_5 | x_4, R = 00110) = p(x_5 | x_4, R_4 = R_5 = 1) \equiv p(x_5 | x_4, R \geq 00011).
\]
Both  $p(x_1, x_2 | x_3, x_4, R \geq 11110)$ and $ p(x_5 | x_4, R \geq 00011)$  can be identified from the observed data, so the imputation model is identifiable. In NP-MMG, imputation submodels are estimated using nonparametric methods such as the k-nearest neighbor or kernel smoothing. Once fitted, these submodels can be used to impute the corresponding missing variables.

\end{example}




\subsection{Selection of the Graph} \label{sec::LG}
MMG fundamentally relies on an undirected graph 
$G$, which encodes conditional dependencies among missing variables and guides the imputation process.
Ideally, we want to choose this graph reflecting our knowledge about the missing variables.
When there is no prior knowledge about the dependency among variables, a common strategy is to estimate $G$ from the observed data, for instance, using complete cases. 


An appealing property of MMG is that when the joint distribution is graphical and faithful to a graph 
$G$ and the missingness mechanism is MCAR, the same graph 
$G$ can be validly used for MMG (see Theorem~\ref{thm: recovery underMCAR}). This result provides justification for data-driven approaches when prior knowledge of the graph is unavailable: one may estimate a sparse undirected graph from the complete data and use it as a plausible working graph. Popular estimation techniques include the graphical lasso \citep{friedman2008sparse} and sparse parallel regression \citep{meinshausen2006high}.

Nevertheless, an estimated graph should only be regarded as a reasonable choice for the MMG's graph but not the true graph for MMG, since correctness is not guaranteed outside the graphical-MCAR setting. In addition, MMG is a family of MNAR assumptions that does not cover every missing data model. Therefore, even if a graph $G$ is perfectly fitted to the complete data, this does not imply that MMG with $G$ is the true model for missing data (in fact, it is impossible to know the true model for missing data without additional information). 
Although the graph in MMG influences the results of downstream analysis, our sensitivity analysis in Section~\ref{sec::sensitivity} shows that MMG-based inferences remain stable under mild perturbations of the graph, suggesting that the method is robust to reasonable choices of $G$.


\section{Statistical Learning with MMG and PAI} 
\label{sec::learning}
While the NP-MMG offers a flexible and model-free approach, it suffers from the curse of dimensionality and can be computationally infeasible. This limitation motivates our consideration of a statistical learning approaches that preserve the principles of MMG while being computationally tractable. In this section, we develop a statistical learning framework under MMG and PAI. We introduce the MMG imputation risk and describe the risk minimization framework. We illustrate the framework with specific examples such as the Gaussian MMG (G-MMG), the Ising MMG (I-MMG), and the Mixture of Product MMG (MP-MMG). 

\subsection{Imputation Risk}
In MMG, the imputation is localized to graph neighborhoods and decomposes into independent submodels corresponding to the connected components of missing entries. For any response vector $r$, 
equation \eqref{eq::MMG1}
shows that its imputation model can be decomposed by to submodels indexed by connected components of $\bar r$ inside $G$. 
Let $s\leq \bar r$ be a response vector of a connected component. We refer to such $s$ as a \textit{connected pattern}. The collection of imputation models  $$
\mathcal{M}=\{
p(x_s | x_{N_G(s)}, R=r): s\leq \bar {r}, s\text{ is a connected component of } G, r\in\{0,1\}^d\}
$$
fully characterize all imputation models under MMG.

We parameterize each model in $\mathcal{M}$ by $\theta_{r,s}$, \ie, $p(x_s | x_{N_G(s)}, R = r;\theta_{r,s}) = p_{\theta_{r,s}}(x_s | x_{N_G(s)}, R=r)$. This parameterization reduces the estimation of  imputation to learning the parameter ${\theta_{r,s}}$ from the data. Since the model $p_{\theta_{r,s}}$ depends on $(x_s, x_{N_G(s)})$, we define a loss function $\mathcal{L}( \theta_{r,s}|x_s, x_{N_G(s)})$ such that $\theta_{r,s}$ can be learned by minimizing the loss. In parametric settings, a common choice for the loss function is the negative log-likelihood.

A powerful feature of PAI supplementing the MMG is that if the same connected component $s$ appears in two response patterns $r_1,r_2$, we have
$p(x_s | x_{N_G(s)}, R=r_1)=p(x_s | x_{N_G(s)}, R=r_2)$;
see Proposition \ref{prop1} for more details. 
This implies that 
$
\theta_{r,s}\equiv\theta_s
$
only needs to be indexed by the connected pattern $s$. 
Therefore, the loss function becomes 
\begin{equation}
\mathcal{L}( \theta_{r,s}|x_s, x_{N_G(s)}) = \mathcal{L}( \theta_{s}|x_s, x_{N_G(s)}),
\label{eq::pai1}
\end{equation}
which greatly reduces the model complexity. 

Based on equation \eqref{eq::pai1} and using the PAI, we define the \textit{MMG imputation risk} as the expected loss over observations where all relevant variables are observed:
\begin{equation}
   \mathcal{R}(\theta_s) = \mathbb{E}[\mathcal{L}( \theta_s|x_s, x_{N_G(s)})\cdot I(R_{\bar{N}_G(s)}=1 )]. 
   \label{eq::risk}
\end{equation}
Accordingly, the population target parameter $\theta_s^*$ is defined as the minimizer of the MMG imputation risk
\[
\theta_s^* = \arg\min_{\theta_s \in \Theta} \mathbb{E}\left[\mathcal{L}(\theta_s|x_s,x_{N_G(s)}) \cdot I(R_{\bar{N}_G(s)}=1 )\right],
\]
where $\Theta$ is the parameter space. 

To estimate $\theta^*$, we construct the empirical MMG imputation risk using all observations where all relevant variables are observed:
\begin{equation}
\hat{\mathcal{R}}_n(\theta_s) = \frac{1}{n}\sum_{i=1}^n \mathcal{L}( \theta_s|\bX_{i,s}, \bX_{i,N_G(s)})\cdot I(\bR_{i,{\bar{N}_G(s)}}=1).
\label{eq::erm}
\end{equation}
The corresponding estimator $\hat{\theta}_s$ is obtained via empirical risk minimization (ERM):
\begin{equation}
\hat{\theta}_s = \arg\min_{\theta_s\in \Theta}  \hat{\mathcal{R}}_n(\theta_s).
\label{eq::estimator}
\end{equation}

This risk minimization framework provides a principled approach to learning imputation models within the MMG paradigm. We refer to it as MMG imputation risk minimization under PAI, or IRM (imputation risk minimization) for short. It is essentially an ERM problem, where the MMG imputation risk corresponds to a localized ERM over graph neighborhoods.
The framework supports a wide range of modeling choices, including likelihood-based models and more flexible machine learning methods, as long as an appropriate loss function is specified. In what follows, we illustrate with the likelihood-based learning, as it is a natural and interpretable starting point for understanding the framework.

\textbf{Likelihood-based learning.} 
Under the likelihood framework, we take the loss function to be the negative log-likelihood: \[\mathcal{L}(\theta_s|x_s, x_{N_G(s)}) = -\log p(x_s, x_{N_G(s)}; \theta_s).\] 
In this case, the IRM estimator in equation \eqref{eq::estimator}
is essentially the maximum likelihood estimator (MLE),
so the population target parameter $\theta_s^*$  satisfies the following properties:

\begin{itemize}
    \item The parameter $\theta_s^*$ minimizes the Kullback-Leibler (KL) divergence between the true conditional distribution and the parametric model:
    \[p(x_s, x_{N_G(s)};\theta_s^*) = \arg\min_{q\in\mathcal{Q}_s} \text{KL}(q||p(x_s, x_{N_G(s)}|R_{\bar{N}_G(s)}=1),\]
    where $\mathcal{Q}_s = \{(x_s, x_{N_G(s)};\theta_s): \theta_s \in \Theta\}$ is the collection of all models in the family.
    \item When the model is correctly specified, \ie, there exists $\Tilde{\theta}_s\in \Theta$ such that $p(x_s, x_{N_G(s)};\Tilde{\theta}_s) = p(x_s, x_{N_G(s)}|R_{\bar{N}_G(s)}=1)$, then $\theta_s^* = \Tilde{\theta}_s$ and the conditional distributions also match \[
p(x_s| x_{N_G(s)};\Tilde{\theta}_s) = p(x_s| x_{N_G(s)},R_{\bar{N}_G(s)}=1).
\]
\end{itemize}

\subsection{Statistical Learning with Gaussian Models: Gaussian MMG (G-MMG)}
The likelihood-based learning framework described above applies to general parametric specifications. To illustrate in a concrete setting, we now focus on the multivariate Gaussian model, which is a natural choice when all variables are continuous.
We begin with a simple example and then formalize the Gaussian MMG (G-MMG) approach.

\begin{example}
\label{example2}
Consider a simple case with $d = 3$ variables and the graph structure is $X_1-X_2-X_3$. There are $2^3 = 8$ possible response patterns in $\{0,1\}^3$. The set of connected patterns includes a total of 6 distinct patterns $\{100, 010, 001, 110, 011, 111\}$, as connected patterns correspond to connected components of missing variables. Each connected pattern defines an imputation submodel indexed by the variables to be imputed. For example, given a response pattern $R = 010$ where $X_2$ is observed, the missing variables $X_1$ and $X_3$ form two disconnected components, yielding two connected patterns: $100$ and $001$. The imputation model under MMG is decomposed into
\[
p(x_1, x_3|x_2, R = 010) = 
p(x_1|x_2, R = 010)
p(x_3|x_2, R = 010).
\]
Now suppose we specify a Gaussian model for each imputation submodel: $p(x_1|x_2, R = 010)=p(x_1|x_2; \theta_{100})$ and $p(x_3|x_2, R = 010)=p(x_3|x_2; \theta_{001})$. The parameter $\theta_{100}=(\mu_{100}, \Sigma_{100})$ characterizes a bivariate Gaussian distribution over $(X_1,X_2)$: \[
\begin{pmatrix}
    X_1\\X_2
\end{pmatrix}\Bigg|R=100 \sim N(\mu_{100}, \Sigma_{100}).
\]
Following PAI, we estimate $\theta_{100}$ by fitting the model to all cases where $R\geq 110$ (variables $X_1$ and $X_2$ are observed). The log-likelihood for $\theta_{100}$ (up to an additive constant)  is 
\[
\ell(\theta_{100}|X_1,X_2) = -\frac{1}{2}\log \det \Sigma_{100} - \frac{1}{2}(X_{1:2}-\mu_{100})^T\Sigma_{100}^{-1}(X_{1:2}-\mu_{100}).
\]
Maximizing this over the available data with $R\geq 110$ yields the MLE $\hat{\theta}_{100} = (\hat{\mu}_{100}, \hat{\Sigma}_{100})$:
\[
\hat{\mu}_{100,j} = \frac{\sum_{i=1}^n \bX_{i,j}I(\bR_i\geq 110)}{\sum_{i=1}^n I(R_i\geq 110)}\quad j=1,2,\]
\[\quad \hat{\Sigma}_{100} = \frac{\sum_{i=1}^n(\bX_{i,1:2}-\hat{\mu}_{100})(\bX_{i,1:2}-\hat{\mu}_{100})^TI(\bR_i\geq 110)}{\sum_{i=1}^n I(\bR_i\geq 110)}.
\]
We compute the conditional distribution from the fitted joint Gaussian
\[
p(x_1|x_2; \hat{\theta}_{100})= N\left( 
\hat{\mu}_{100,1} + \hat{\Sigma}_{12} \hat{\Sigma}_{22}^{-1}(x_2 - \hat{\mu}_{100,2}),\ 
\hat{\Sigma}_{11} - \hat{\Sigma}_{12} \hat{\Sigma}_{22}^{-1} \hat{\Sigma}_{21}
\right)
\]
and impute $X_1$ accordingly.
Similarly, we estimate $\theta_{001}$ by fitting a bivariate Gaussian model to data with $R\geq 011$ (\ie, both $X_2$ and $X_3$ are observed) and derive $p(x_3|x_2; \hat{\theta}_{001})$ to impute $X_3$. 
We refer to pattern $110, 011$ as \emph{model patterns}, as they are patterns derived under MMG and PAI and are used for estimating the parameters of the imputation submodels. See Appendix~\ref{appendix: model pattern} for further discussion of connected patterns and model patterns.
\end{example}

This example motivates the \textit{Gaussian MMG (G-MMG)} approach, where all variables are continuous and imputation submodels are constructed using multivariate Gaussian distributions over the relevant variables. Formally,  each imputation submodel
\( p(x_{s} | x_{N_G(s)}, R = r) \) 
is based on the joint Gaussian distribution of \( X_{\bar{N}_G(s)} \):
\[
X_{\bar{N}_G(s)} \mid R = r \sim \mathcal{N}(\mu_{s}, \Sigma_{s}),
\]
where the parameters \( \mu_{s}, \Sigma_{s} \) are the same as the following marginal Gaussian model 
\[
X_{\bar{N}_G(s)} \mid R_{\bar{N}_G(s)} = 1 \sim \mathcal{N}(\mu_{s}, \Sigma_{s}).
\]
G-MMG is a special case of the IRM, as we can easily write down the log-likelihood and the MLE. For each connected pattern $s$, the log-likelihood of the associated parameter $\theta_{s}=(\mu_{s}, \Sigma_{s})$ is (up to an additive constant)
\[
\ell(\theta_{s}|X_{\bar{N}_G(s)}) = -\frac{1}{2}\log \det \Sigma_{s} - \frac{1}{2}(X_{\bar{N}_G(s)}-\mu_{s})^T\Sigma_{s}^{-1}(X_{\bar{N}_G(s)}-\mu_{s}).
\]
In short, G-MMG adopts a marginal modeling strategy: we first fit a Gaussian model to observations with $R_{\bar{N}_G(s)}=1$, and then derive the conditional distribution $ p(x_{s} | x_{N_G(s)}, R = r)$ from this marginal model for imputation. 


\begin{remark}
Under the Gaussian distribution, the conditional distributions needed for imputation are easy to derive. Importantly, we do not require that $X_{\bar{N}_G(s)} \mid R_{\bar{N}_G(s)}=1$ is indeed Gaussian. Instead, we only borrow the parameters (mean and covariance) from the fitted Gaussian model.  
The parametric models in this framework are constructed marginally (e.g., via Gaussian models for subsets of variables) and then transformed into conditional distributions for imputation. However, this approach is flexible and accommodates multiple variants: the mean function can be arbitrarily specified under a conditional Gaussian model, and one can use other flexible parametric models for the conditional distribution.  
\end{remark}

\subsection{Statistical Learning with Ising Models: Ising MMG (I-MMG)}
When all variables are binary, a popular approach is to model the joint distribution using an Ising model.  
MMG builds imputation models based on Ising models. I-MMG assumes that each imputation submodel $p(x_{s} | x_{N_G(s)}, R = r)$ is the conditional distribution derived from fitting the following marginal model:
\[
X_{\bar{N}_G(s)} | R_{\bar{N}_G(s)} = 1 \sim \text{Ising}(\theta_{s}, \Lambda_{s}).
\]
I-MMG is a special case of the IRM, as we can express the log-likelihood as
\[
\ell(\theta_{s}, \Lambda_{s}|X_{\bar{N}_G(s)}) 
= \theta_{s}^T X_{\bar{N}_G(s)} 
+ X_{\bar{N}_G(s)}^\top \Lambda_{s} X_{\bar{N}_G(s)} 
- \log Z(\theta_{s},\Lambda_{s}),
\]
where $Z(\theta_{s},\Lambda_{s})$
is the partition function (normalizing factor) to ensure that $p(x_{\bar{N}_G(s)}; \theta_{s})$ is a valid PMF.

\begin{example}
\label{example3}
We revisit Example \ref{example1} for I-MMG. Given a response pattern $R=00110$, the connected patterns are 11000 and 00001. The I-MMG assumes that the imputation submodel $p(x_5 | x_4, R=00110)$ is the conditional PMF from fitting the following the Ising model $\text{Ising }(\theta_{00001}, \Lambda_{00001})$ to cases with $R\geq 00011$. Similarly, the other imputation submodel $p(x_1, x_2 | x_3, x_4, R = 00110)$ is assumed to be the conditional PMF from fitting $\text{Ising }(\theta_{11000}, \Lambda_{11000})$ to cases with $R\geq 11110$. 
\end{example}

Note that G-MMG and I-MMG are particularly interesting models for statistical learning because they are also graphical models under appropriate assumptions. 
Under MCAR, the graph in the graphical model
can be used as the graph for MMG; see Theorem~\ref{thm: recovery underMCAR} for more details.

\subsection{Statistical Learning for Mixed-type Data: Mixture of Product MMG (MP-MMG)}

When the variables are
of mixed types (e.g., some are continuous, some are categorical), it is not easy to impose a parametric model.  
A common strategy is to use a mixture of product (MP; \citealp{suen2023modeling}) model, which assumes that
\begin{equation}
    p(x) = \sum_{k=1}^{K} \pi_k \prod_{j=1}^{d} p(x_j; \theta_{k,j}),
    \label{eq: MP}
\end{equation}
where $p(x_j; \theta_{k,j})$ is a univariate model of variable $x_j$ parameterized by $\theta_{k,j}$ and $\pi_k \geq 0$ with $\sum_k \pi_k = 1$.  We use the abbreviation $X \sim \text{MP}(\pi, \theta)$ to refer to the model in equation (\ref{eq: MP}). The reasoning behind is that modeling an individual variable using a parametric form is easy, but modeling the dependency among variables is hard. Thus, equation (\ref{eq: MP}) uses the mixture form to handle the dependency.  
When there is no missing data, estimating the parameters $\pi$ and $\theta$ is a standard problem in mixture models and can be done using the Expectation-Maximization (EM) algorithm.

Similar to the Gaussian and Ising models, the MP models can be combined with the MMG framework, leading to MP-MMG. MP-MMG assumes that the imputation submodel $p(x_{s} | x_{N_G(s)}, R = r)$ is the conditional distribution derived from fitting the marginal model
\[
X_{\bar{N}_G(s)} \mid R_{\bar{N}_G(s)} = 1 \sim \text{MP}(\pi_{s}, \theta_{s}).
\]
MP-MMG is also a special case of IRM, as the log-likelihood of $(\pi_{s}, \theta_{s})$ is
\[
\ell(\pi_{s}, \theta_{s}|X_{\bar{N}_G(s)}) 
= \log\Bigg(
\sum_{k=1}^{K}\pi_{s,k}\prod_{j \in \bar{N}_G(s)} p(x_{j}; \theta_{s,k,j})
\Bigg).
\]

\begin{example}
\label{example4}
We reconsider Example \ref{example1}. Given the pattern $R=00110$, the MP-MMG assumes that the imputation submodel $p(x_1, x_2 | x_3, x_4, R = 00110)$ is the conditional distribution derived from fitting the model $X_1,\dots,X_4 \sim \text{MP}(\pi_{11000}, \Lambda_{11000})$ to cases with $R\geq 11110$. Similarly, $p(x_5 | x_4, R=00110)$ is derived from fitting model $X_4,X_5 \sim \text{MP}(\pi_{00001}, \Lambda_{00001})$ to cases with $R\geq 00011$. The parameters $(\theta_{11000}, \Lambda_{11000})$ and 
$(\theta_{00001}, \Lambda_{00001})$ can be estimated using the EM algorithm.
\end{example}

\section{Statistical Theories of MMG} \label{sec::theory}
In this section, we establish theoretical properties of MMG (and PAI).

\subsection{Connections to CCMV}
We build a link between MMG-PAI and the well-known complete-case missing value (CCMV) assumption \citep{little1993pattern}, which essentially assumes that  $p(X_{\bar{r}}|X_r,R=r)=p(X_{\bar{r}}|X_r,R=1)$ for every pattern $R=r$.

\begin{theorem}
\label{thm: PAI_CCMV}
If the Markov missing graph $G=(V,E)$ is fully connected, then MMG and PAI coincides with CCMV.
\end{theorem}

Theorem~\ref{thm: PAI_CCMV} follows immediately by noticing that $\bar{N}_G(s)=V$ and $N_G(s) = V\setminus s$ when $G$ is fully connected. PAI thus becomes $p(x_s|x_{V\setminus s},R=r) = p(x_s|x_{V\setminus s}, R_V=1)$, which is exactly CCMV. 
Note that a fully connected MMG does not provide any reduction
in the construction of imputation models,
so we always have to use all variables for imputing missing variables.

It is of interest to explore how CCMV connects to PAI and MMG  with a non-fully connected graphical structure. 
To study this, we consider the monotone missing data scenario, i.e., the set of response patterns can be expressed as $\{\, \underbrace{11\cdots 1}_{l} \underbrace{00\cdots 0}_{d-l} : l\in[d] \, \}$. Under monotonicity, each pattern is uniquely determined by $l$, the number of observed variables. Let $r^{(l)}$ be the response pattern with exactly $l$ observed variables.




\begin{theorem}
\label{thm: PAI_CCMV_monotone}
Suppose complete cases follow a graphical model faithful to $G$ that has a chain structure $X_1-X_2-\cdots-X_d$. Assume monotone missingness. Then CCMV is equivalent to PAI and MMG with respect to $G$, i.e.,
\[
p(x_{l+1},\dots,x_d|x_1,\dots,x_l,R=r^{(l)})
=p(x_{l+1},\dots,x_d|x_l,R=1).
\]
\end{theorem}

Theorem~\ref{thm: PAI_CCMV_monotone} suggests that under monotone missing data, if the complete case is faithful to a graph $G$ with a chain structure, then it is equivalent to the MMG (with PAI) with respect to the same graph $G$.

\begin{example}
\label{example: PAI_CCMV_monotone}
Let $G$ be the chain $X_1-X_2-X_3-X_4-X_5$ and suppose the missingness is monotone in an order of $X_1 \prec X_2 \prec \cdots \prec X_5$. Consider the response pattern $R=11100$ where variables $X_4,X_5$ are missing. Under PAI and MMG, the imputation model is  $p(x_4,x_5|x_3,R=11000)=p(x_4,x_5|x_3, R\geq 00111)$. Monotonicity implies that $R\geq 00111$ reduces to $R=11111$, so in this case PAI uses only the complete cases:
$p(x_4,x_5|x_3,R=11100) = p(x_4,x_5|x_3,R=11111)$.
The CCMV assumption yields
$p(x_4,x_5|x_1,x_2,x_3,R=11100) = p(x_4,x_5|x_1,x_2,x_3,R=11111)$. When the complete cases follow a graphical model faithful to $G$, we have $p(x_4,x_5|x_1,x_2,x_3,R=11111)=p(x_4,x_5|x_3,R=11111)$. Therefore, the imputation models coincide.
\end{example}


\subsection{Recovery under MCAR}
We now establish a fundamental recovery guarantee for MMG under missing completely at random (MCAR). This recovery property ensures that the imputation  can reproduce the full-data distribution when missingness is independent of the data. The following theorem formalizes this property.

\begin{theorem}
If the missingness mechanism is MCAR and the underlying data distribution is faithful to a graph $G$, then NP-MMG with respect to $G$ recovers the true model. Moreover, if the distribution of $X$ follows a multivariate normal, then G-MMG is asymptotically more efficient than the complete-case analysis. 
\label{thm: recovery underMCAR}
\end{theorem} 
A formal proof is provided in Appendix~\ref{appendix: proof}. 
Theorem \ref{thm: recovery underMCAR} shows that
when the joint distribution is graphical to a graph $G$
and the missingness is MCAR, the same graph can be used in the MMG. In MCAR, a complete case analysis yields an unbiased estimate of the model. This result suggests that we may use complete data to find a graph quantifying the dependency and use it for MMG.

The same efficiency gain in Theorem~\ref{thm: recovery underMCAR} also holds for I-MMG since Ising model is also a graphical model. 
While the proof is given for G-MMG, we would expect analogous efficiency gains for other MMG variants, as MMG use more information than the complete-case analysis.

\subsection{Efficiency Theory}  \label{sec::efficiency}
We now develop the efficiency theory for MMG. 
We focus on NP-MMG and consider estimating the mean of a variable for simplicity. Without loss of generality, we take the target parameter to be $\mathbb{E}(X_1)$. We can express $\mathbb{E}(X_1)$ as
\begin{equation}
\mathbb{E}(X_1) = \sum_r \mathbb{E}(X_1 I(R = r)) = \sum_{r: r_1 = 1} \mathbb{E}(X_1 I(R = r)) + \sum_{r: r_1 = 0} \mathbb{E}(X_1 I(R = r)),
\label{eq: EX1}
\end{equation}
where $I(\cdot)$ is the indicator function. The first term is directly identifiable from the observed data, leaving only the second term $\sum_{r: r_1 = 0} \mathbb{E}(X_1 I(R = r))$ to be identified.

Let $s_1, \cdots, s_J$ denote the response vector of connected subsets of the graph $G$ such that $X_1 \in s_j$ for all $j$. That is, each $s_j$ is a connected subset of $\{X_1,\cdots, X_n\}$ containing $X_1$. Note that $s_j$ includes the connected component of $G$ containing $X_1$ but also includes subsets of this connected component as long as the subset contains $X_1$. 
Note that while $s_j$ is a connected subset of $G$, 
it may be a connected component for some response vector $r$ (or $\bar r$).

For each subset $s_j$, we define the collection of response patterns 
\begin{equation}
    \mathbb{S}_j = \{ r : r_{s_j} = 0, r_{N_G(s_j)} = 1 \},
    \label{eq::c1}
\end{equation}
where $r_{s_j}=0$ means that all variables in $s_j$ are missing, and $r_{N_G(s_j)} = 1$ means that all variables in $N_G(s_j)$ are observed. Thus, the collection $\mathbb{S}_j$ represents patterns whose information on $X_1$ is from the boundary $N_G(s_j)$.

For any pattern $r$ with $r_1 = 1$, we define the mapping $\psi(r)\in\{0,1\}^d$ to be the response vector of the connected component of $r$ containing $X_1$. Clearly, $\psi(r) \in \{s_1, \cdots, s_J\}$. 
The event $\psi(\bar{R})=s_j$ refers to cases where $X_1$ is missing and the missing variables in $R$ contains a connected component $s_j$ that includes $X_1$. With this, we can rewrite $\mathbb{S}_j$ in equation \eqref{eq::c1} as 
\begin{equation}
\mathbb{S}_j = \{ r : \psi(\bar{r})= s_j \}.
\label{eq::c2}
\end{equation}
Equation \eqref{eq::c2} provides an alternative way to interpret $\mathbb{S}_j$: $\mathbb{S}_j$ is the collection
of patterns where $X_1$ is missing
and equation \eqref{eq::c1} indicates that we can recover the missing information of $X_1$
using the boundary of $s_j$.

With this notation and property, we show how the term in Equation~\eqref{eq: EX1} is identified and estimated by using the regression adjustment approach and inverse probability weighting (IPW). 

\subsubsection{Regression Adjustment}
We begin with the regression adjustment approach, which identifies Equation~\eqref{eq: EX1} by expressing 
$\sum_{r: r_1 = 0} \mathbb{E}(X_1 I(R = r))$ 
as an average of conditional expectations. We denote 
$\mu_r = \mathbb{E}(X_1 I(R = r))$, where $r_1=0$.
Because $r_1 = 0$ implies $\bar{r}_1 = 1$, the function \( \psi \) can be used to identify the connected subset containing \( X_1 \) under the pattern $r$. Suppose that \( \psi(\bar{r}) = s_j\). Let \( x_{-1} \) denote the vector \( (x_2, \cdots, x_d) \). Under MMG and PAI, we  express $\mu_r$ as
\begin{equation}
\label{eq: regression adjustment}
\mu_r = \mathbb{E}(X_1 I(R = r)) = \mathbb{E}(m_{s_j}(X_{N_G(s_j)}) I(R = r))
\end{equation}
where the regression function $m_{s_j}$ is 
\[
m_{s_j}(x_{N_G(s_j)})=\int x_1 p(x_1 | x_{N_G(s_j)}, R \geq \bar{N}_G(s_j)) dx_1.
\]
A complete derivation is provided in Appendix~\ref{sec::RA}. The regression function is clearly identifiable. Since \( m_{s_j}(X_{N_G(s_j)}) \) is the same across all patterns \( r \) with \( \psi(\bar{r}) = s_j \), 
we aggregate over patterns and rewrite the summation as
\[
\sum_{r: r_1 = 0} \mathbb{E}(X_1 I(R = r)) 
= \sum_{j=1}^J \mathbb{E}(m_{s_j}(X_{N_G(s_j)}) I(\psi(\bar{R}) = s_j)),
\]
which leads to
\begin{equation}
\mu = \sum_{r: r_1 = 1} \mathbb{E}(X_1 I(R = r)) + \sum_{j=1}^J \mathbb{E}(m_{s_j}(X_{N_G(s_j)}) I(\psi(\bar{R}) = s_j)). 
\label{eq: regression adjustment2}
\end{equation}
Equation~\eqref{eq: regression adjustment2} shows that we can identify $\mu$ using the regression adjustment. A corresponding regression adjustment estimator is 
\begin{equation*}
    \hat{\mu}_{\text{ra}}=\sum_{r: r_1 = 1} \Bigg(\frac{1}{n}\sum_{i=1}^n \bX_{1i}I(\bR_i=r)\Bigg)
    + \sum_{j=1}^J \Bigg(\frac{1}{n}\sum_{i=1}^n 
    \hat{m}_{s_j}(\bX_{i,N_G(s_j)}) I(\psi(\bar{\bR}_{i}) = s_j))
    \Bigg).
\end{equation*}

\subsubsection{Inverse Probability Weighting}
We next turn to the inverse probability weighting approach. Consider a response pattern $r$ such that $\psi(\bar{r}) = s_j$. From the previous result, we have
\begin{equation*}
\mu_r = \mathbb{E}(X_1 I(R = r)) = \mathbb{E}(m_{s_j}(X_{N_G(s_j)}) I(R = r)).
\end{equation*}
Another calculation shows that
\begin{equation}
\mu_r = \mathbb{E}(X_1 I(R = r)) = \mathbb{E}(X_1 \cdot \mathcal{O}_r(X_{N_G(s_j)}) \cdot I(R \ge \bar{N}_G(s_j))).
\label{eq: IPW}
\end{equation}
where the odds is defined as $\mathcal{O}_r(X_{N_G(s_j)}) = \frac{P(R = r | x_{N_G(s_j)})}{P(R \ge \bar{N}_G(s_j) | x_{N_G(s_j)})}$, which is identifiable because $x_{N_G(s_j)}$ is observed in both $R=r$ and $R\geq \bar N_G(s_j)$.
A full derivation of Equation \eqref{eq: IPW} is provided in Appendix \ref{sec::IPW}. 

Equation~\eqref{eq: IPW} reveals an interesting feature: the dependency on the pattern $r$ is fully captured by $\mathcal{O}_r$, while the other parts remain the same for any $r$ satisfying $\psi(\bar{r}) = s_j$. This feature suggests that we can aggregate expectations across such patterns by defining
\begin{align*}
\mu_{s_j} &\equiv \sum_{r: \psi(\bar{r}) = s_j} \mathbb{E}(X_1 I(R = r)) \\
&= \mathbb{E} \left( X_1 \cdot \sum_{r: \psi(\bar{r}) = s_j} \mathcal{O}_r(X_{N_G(s_j)}) \cdot I(R \ge \bar{N}_G(s_j)) \right) \\
&= \mathbb{E} \left( X_1 \cdot \mathcal{O}_{s_j}(X_{N_G(s_j)}) \cdot I(R \ge \bar{N}_G(s_j)) \right),
\end{align*}
where
\begin{equation*}
\mathcal{O}_{s_j}(X_{N_G(s_j)}) = \frac{p(\psi(\bar{R}) = s_j | X_{N_G(s_j)})}{p(R \ge \bar{N}_G(s_j) | X_{N_G(s_j)})} 
= \frac{P(R \in \mathbb{S}_j | X_{N_G(s_j)})}{P(R \ge \bar{N}_G(s_j) | X_{N_G(s_j)})}
\end{equation*}
represents the odds of observing any patterns whose missing variables form the connected subset $s_j$, relative to the patterns where all variables in $\bar{N}_G(s_j)$ being observed, conditional on $X_{N_G(s_j)}$. 

Therefore, $\mu$ can be identified using IPW:
\begin{equation*}
\mu = \sum_{r: r_1 = 1} \mathbb{E}(X_1 I(R = r)) + \sum_{j=1}^J \mathbb{E}(X_1 \cdot \mathcal{O}_{s_j}(X_{N_G(s_j)}) \cdot I(R \ge \bar{N}_G(s_j))).
\end{equation*}
The IPW estimator for $\mu$ is given by
\begin{equation*}
    \hat{\mu}_{\text{IPW}}=\sum_{r: r_1 = 1} \Bigg(\frac{1}{n}\sum_{i=1}^n \bX_{1i}I(\bR_i=r)\Bigg)
    + \sum_{j=1}^J \Bigg(\frac{1}{n}\sum_{i=1}^n 
    \bX_{1i}\hat{\mathcal{O}}_{s_j}(\bX_{i, N_G(s_j)})
     I(\bR_i \ge \bar{N}_G(s_j))
    \Bigg).
\end{equation*}

\subsubsection{Efficient Influence Function}
We have shown that 
$\mu_r = \mathbb{E}(X_1 \mathbb{I}(R = r))$ with $r_1 = 0$ can be aggregated into
\[
\mu_{s_j} = \sum_{\psi(\bar{r}) = s_j} \mathbb{E}(X_1 I(R = r)) = \mathbb{E}(X_1 I(\psi(\bar{R}) = s_j)).
\]
This aggregation motivates us to study the linear form in the efficient influence function (EIF) of $\mu_{s_j}$.

\begin{theorem}
The linear form of the EIF of $\mu_{s_j}$ is

\begin{align*}
\mathrm{LEIF}_{s_j}(X_1, X_{N_G(s_j)}, R)
&= X_1 \cdot \mathcal{O}_{s_j}(X_{N_G(s_j)}) \cdot  I(R \ge \bar{N}_G(s_j)) 
\\
&\quad +m_{s_j}(X_{N_G(s_j)})\Bigg(I(\psi(\bar{R}) = s_j) - I(R \ge \bar{N}_G(s_j))\cdot \mathcal{O}_{s_j}(X_{N_G(s_j)}) 
\Bigg).
\end{align*}
The EIF of $\mu_{s_j}$ is thus
\[
\mathrm{EIF}_{s_j}(X_1, X_{N_G(s_j)}, R)=\mathrm{LEIF}_{s_j}(X_1, X_{N_G(s_j)}, R) - \mathbb{E}[X_1 I(\psi(\bar{R}) = s_j)].
\]
\label{thm: EIF}
\end{theorem}

The linear form in Theorem~\ref{thm: EIF} has two interpretations. The first way is by expressing it as
\begin{equation*}
\begin{aligned}
\text{LEIF}_{s_j}(X_1, X_{N_G(s_j)}, R) 
&= \underbrace{m_{s_j}(X_{N_G(s_j)}) I(\psi(\bar{R}) = s_j)}_{\text{Regression adjustment}} \\
&\quad + \underbrace{[X_1 - m_{s_j}(X_{N_G(s_j)})] \mathcal{O}_{s_j}(X_{N_G(s_j)}) I(R \geq \bar{N}_G(s_j))}_{\text{Augmentation}}.
\end{aligned}
\end{equation*}
This expression shows that the LEIF is essentially a regression adjustment method with an augmentation to improve the efficiency. Using the conditional mean, the augmentation term has mean 0, that is,
\[
\mathbb{E} \left\{ \left[X_1 - m_{s_j}(X_{N_G(s_j)}) \right] \mathcal{O}_{s_j}(X_{N_G(s_j)})I(R \geq \bar{N}_G(s_j)) \right\} = 0.
\]
The second approach is to express the linear term as an IPW with augmentation, namely,
\begin{equation*}
\begin{aligned}
\text{LEIF}_{s_j}(X_1, X_{N_G(s_j)}, R) 
&= \underbrace{X_1 \mathcal{O}_{s_j}(X_{N_G(s_j)}) I(R \geq \bar{N}_G(s_j))}_{\text{IPW}} \\
&\quad + \underbrace{m_{s_j}(X_{N_G(s_j)}) \left[I(\psi(\bar{R}) = s_j) - \mathcal{O}_{s_j}(X_{N_G(s_j)}) I(R \geq \bar{N}_G(s_j)) \right]}_{\text{Augmentation}}.    
\end{aligned}
\end{equation*}
This shows that the LEIF can be viewed as an augmented IPW form. It can be shown that the augmentation term has mean 0:
\[
\mathbb{E} \left\{ m_{s_j}(X_{N_G(s_j)}) \left[I(\psi(\bar{R}) = s_j) - \mathcal{O}_{s_j}(X_{N_G(s_j)}) I(R \geq \bar{N}_G(s_j)) \right] \right\} = 0.
\]
Based on Theorem~\ref{thm: EIF}, an efficient estimator for $\mu_{s_j}$ is 
\begin{equation*}
\begin{aligned}
 \hat\mu_{s_j} &= \frac{1}{n}\sum_{i=1}^n \Bigg[
  \bX_{i,1} \cdot \hat{\mathcal{O}}_{s_j}(\bX_{i,N_G(s_j)}) \cdot  I(\bR_i \ge \bar{N}_G(s_j)) 
\\
&\quad + \hat{m}_{s_j}(\bX_{i,N_G(s_j)})\Bigg(I(\psi(\bar{\bR}_i) = s_j) - I(\bR_i \ge \bar{N}_G(s_j))\cdot \hat{\mathcal{O}}_{s_j}(\bX_{i,N_G(s_j)}) 
\Bigg) \Bigg] .
\end{aligned}
\end{equation*}


Thus, the corresponding efficient  estimator for $\mu=\mathbb{E}[X_1]$ is 
\begin{equation*}
\hat{\mu} = \sum_{r:r_1=1}\frac{1}{n}\sum_{i=1}^n X_{1i}I(R_i=r)+\sum_{j=1}^J\hat{\mu}_{s_j}  
\end{equation*}. 
This estimator
has a multiple robustness property under the following assumptions:

\begin{itemize}
\item[(A1)] {\bf Donsker condition.} For each $j\in \{1,\dots,J\}$, $\hat{\mathcal{O}}_{s_j}$ is in a uniformly bounded Donsker class $\mathcal{F}_j$ and $\hat{m}_{s_j}$ is in a uniformly bounded Donsker class $\mathcal{G}_j$. There exist functions $\mathcal{O}^*_{s_j}$ and $m^*_{s_j}$ such that $\norm{\hat{\mathcal{O}}_{s_j} - \mathcal{O}^*_{s_j}}_{L_2(P)}=o_P(1)$ and 
$\norm{\hat{m}_{s_j} - m^*_{s_j}}_{L_2(P)}=o_P(1)$, where $\norm{f}_{L_2(P)}=(\int |f|^2dP)^{1/2}$ is the $L_2(P)$ norm of a function $f$.

\item[(A2)] {\bf Uniformly bounded}. $X_1,  \mathcal{O}_{s_j},m_{s_j}$ are uniformly bounded by a constant $M>0$.

\item[(A3)] {\bf Multiply-robustness.} $\sum_{j=1}^J \norm{\hat{\mathcal{O}}_{s_j} - \mathcal{O}_{s_j}}_{L_2(P)}\norm{\hat{m}_{s_j} - m_{s_j}}_{L_2(P)}=o_P(1)$.

\item[(A4)] {\bf Entropy bound.} $\mathcal{F}_j$ satisfies the uniform entropy bound \citep{van2000asymptotic}: \[\int_{0}^{\infty} \sup_Q\sqrt{\log N(\epsilon\norm{\bar{F}_j}_{Q,2}, \mathcal{F}_j, L^2(Q))}d\epsilon<\infty,
\]
where the supremum is over all discrete probability measures $Q$ on $X$,
$\bar{F}_j$ is the envelope function for $\mathcal{F}_j$ such that $P\bar{F}_j<\infty$ with $P$ being the probability measure for $X$, $N(\epsilon\norm{\bar{F}_j}_{Q,2}, \mathcal{F}_j, L^2(Q))$ is the covering number of the class $\mathcal{F}_j$ with respect to the $L_2(Q)$ norm where $\norm{\bar{F}_j}_{Q,2}=Q\bar{F}_j^2$. 
Similarly, we assume $\mathcal{G}_j$ satisfies the uniform entropy bound.
\end{itemize}

Assumption (A1) is a standard assumption that
the requires our estimators of nuisance parameters are stable. This is a mild assumption as most of the parametric models form a uniformly bounded Donsker class when the parameter space is a compact set (see Example 19.7 of \citealp{van2000asymptotic}). 
(A2) is a mild uniformly bounded assumption to ensure the asymptotic linear form of the efficient influence function is a uniformly bounded function. 
(A3) is the assumption of `multiply-robustness' that we need one of the two models to be consistent. Note that (A1) only requires estimators to have a stable limit but does not require this limit to be the correct odds or regression function. (A3) requires one of the limiting targets to be the correct odds/regression function. (A4) is the uniform entropy bound that ensures the `use data twice' procedure of the efficient estimator is still consistent. It is needed because we use the data for  both estimating the nuisance and computing the efficient estimator. Again, it is a mild assumption that most of the parametric models with compact support satisfy this.

\begin{theorem}[Multiple robustness] 
Under assumptions (A1-4),
the estimator $\hat{\mu}$ is consistent.
\label{thm: multiple robustness}
\end{theorem}
The key multiply-robustness assumption is (A3) that for every pair of models, we need one of them to be a consistent estimator to achieve the consistency of the final estimator. We may obtain asymptotic normality of the estimator $\hat\mu$ by multiplying $\sqrt{n}$ to the left-hand-side of assumption (A3).

\subsection{Graphical Properties of MMG and PAI}    \label{sec::graph}
\subsubsection{Model Reduction due to PAI}

As mentioned in Section \ref{sec::learning},
the PAI reduces the model complexity of the MMG imputation models. Here we formally describe how the structure of the graph $G$ shapes the imputation model under MMG and PAI. Specifically, we show that the imputation submodels are induced locally by the graph. 
\begin{proposition}
Let $s$ be a connected component of $G$. 
Consider two patterns $r_1 \neq r_2$ such that $
r_{1,j} = r_{2,j} =0$ for all $j$ with $s_j=1$ (missingness agreement in  $s$) and 
$r_{1,k} = r_{2,k} = 1$ for all $k \in N_G(s)$ (observed  agreement in the neighborhood).
Then under MMG and PAI, 
\[
p(x_s \mid x_{N_G(s)}, R = r_1) = p(x_s \mid x_{N_G(s)}, R = r_2).
\]
\label{prop1}
\end{proposition}

Proposition \ref{prop1} follows immediately from the PAI.
It means that if a connected subset $s$ is also a connected component in both $r_1$ and $r_2$, then the imputation model for $X_s$ is the same in both patterns.

Importantly, Proposition~\ref{prop1} implies a key computational benefit: PAI reduces the number of unique imputation models needed, as different patterns sharing the same local structure would yield the same imputation submodel.    


\begin{example}
\label{example5}
Consider Example~\ref{example1} and two different missing patterns: case 1: $r_1 = 00110$ (only $X_3, X_4$ observed) and case 2: $r_2 = 10110$ ($X_1, X_3, X_4$ observed). In both cases, variable $X_5$ is missing and its neighbor $X_4$ is observed. The connected pattern of interest is $s=00001$. 
According to MMG, the imputation submodel for $X_5$ in case 1 is $p(x_5 | x_4, R = 00110)$ and in case 2 is $p(x_5 | x_4, R = 10110)$. PAI implies that 
\[
p(x_5 | x_4, R = 00110) = p(x_5 | x_4, R \geq  00011)=p(x_5 | x_4, R = 10110).
\]
Thus, the imputation model for $X_5$ is identical for both patterns, which reflects the statement of Proposition~\ref{prop1}. 

\end{example}

\subsubsection{Nested Submodel}
 MMG and PAI also induce relationships between imputation submodels defined on different sets of variables. 
\begin{proposition}
Let $A \subset \{1, \ldots, d\}$ and $B \subset A$ such that $\bar{N}_G(A) = \bar{N}_G(B)$. Then the imputation submodel under set $A$
\[
p(x_A \mid x_{N_G(A)}, R_A = 0, R_{N_G(A)} = 1)
\]
implies the imputation submodel on subset B from
\begin{align*}
p(x_B \mid x_{N_G(B)}, R_B = 0, R_{N_G(B)} = 1) &= p(x_B \mid x_{N_G(B)}, R_A = 0, R_{N_G(A)} = 1) \\
& = \frac{p(x_A \mid x_{N_G(A)}, R_A = 0, R_{N_G(A)} = 1)}{\int p(x_A \mid x_{N_G(A)}, R_A = 0, R_{N_G(A)} = 1)dx_{B}}.
\end{align*}
\label{prop2}
\end{proposition}
By Proposition \ref{prop2}, the imputation model for a larger set of variables leads to the same imputation
model of a smaller set of variables, provided their neighborhood coincide.

\begin{example}
\label{example6}
Consider Example~\ref{example1}. Let $A = \{X_1,X_2,X_3\}$ and $B = \{X_2,X_3\}$. Since $\bar{N}_G(A) = \bar{N}_G(B) = \{X_1,X_2,X_3,X_4\}$, Proposition \ref{prop2} shows that the imputation submodel $p(x_1,x_2,x_3|x_4,R=0001r_{5})$ implies the imputation submodel $p(x_2,x_3|x_1,x_4,R=1001r_5)$ for any $r_5\in\{0,1\}$ via
\begin{align*}
p(x_2,x_3|x_1,x_4,R=1001r_5) 
&= p(x_2,x_3|x_1,x_4,R\geq11110)\\
& = \frac{p(x_1, x_2,x_3|x_4,R\geq11110)}{\int p(x_1,x_2,x_3|x_4,R\geq11110) dx_2dx_3}\\
& = \frac{p(x_1, x_2,x_3|x_4,R= 0001r_{5})}{\int p(x_1,x_2,x_3|x_4,R=0001r_{5}) dx_2dx_3},
\end{align*}
where we use the MMG and PAI in the first and third equalities. 
\end{example}

\subsubsection{Local Imputation Submodel}
The previous properties show how MMG governs invariance across patterns and nested variable sets. A natural next step is to investigate the role of locality in MMG. In particular, we show that the imputation model for a single connected component depends only on the local
information, namely the neighborhood of the connected component. 
\begin{proposition}
Let $r$ be a response pattern such that $\bar{r}$ forms a single connected component. If the MMG with respect to graphs $G_1$ and $G_2$ satisfies $N_{G_1}(\bar{r}) = N_{G_2}(\bar{r})$, then the imputation model under $G_1$ and $G_2$ coincide, i.e.,
\[
p_{G_1}(x_{\bar{r}} | x_r, R = r) = p_{G_2}(x_{\bar{r}} | x_r, R = r),
\]
where $p_G$ is the imputation model under MMG $G$.
\label{prop3}
\end{proposition}
Proposition \ref{prop3} is a straightforward result from
the definition of MMG (equation \eqref{eq::MMG1}).
Proposition \ref{prop3} suggests that any two MMGs with identical neighborhoods of a connected component will yield the same imputation model, even if their global structure differ. Proposition \ref{prop3} immediately implies the following result on multiple connected components.

\begin{corollary}
Let $r$ be a response pattern such that $\bar{r} = s_1 + \cdots + s_K$ with each $s_k$ being a connected component. Let $G_1$ and $G_2$ be two graphs such that $N_{G_1}(s_k) = N_{G_2}(s_k)$ for all $k$. Then the imputation model implied by the MMGs
\[
p_{G_1}(x_{\bar{r}} | x_r, R = r) = p_{G_2}(x_{\bar{r}} | x_r, R = r).
\]
\end{corollary}
\begin{example}
\label{example7}
Let $G_1$ be the graph in Figure~\ref{fig:example1}, and $G_2$ be a modified graph where edges (1,2) and (2,4) are removed. For the response pattern $r = 11110$, Proposition~\ref{prop3} applies because $\bar{r}$ forms a single connected component $\{X_5\}$ and $N_{G_1}(\{X_5\})=N_{G_2}(\{X_5\})=\{X_4\}$. Thus, both graphs yield the same imputation submodel $p(x_5 | x_4, R = 11110)$. This suggests that the imputation model remains unchanged as long as the changes to the graph do not affect the variable being imputed.    
\end{example}

It is important to note that the requirement on connected components in
Proposition~\ref{prop3} is necessary. A counterexample is as follows.
\begin{example}
\label{example8}
 Consider a scenario with variables $X_1, X_2, X_3, X_4$ and two graphs: $G_1$ has edges $(1,2), (2,3), (3,4), (1,4)$ and $G_2$ has edges $(1,2), (2,3), (3,4)$. Let the pattern $r = 0101$ so that variables $X_1$ and $X_3$ are missing. Clearly, in both graphs the neighbor $N_{G_1}(\bar{r})=N_{G_2}(\bar{r})=\{X_2,X_4\}$.
 However, MMG with respect to $G_1$ implies an imputation model
\[
p_{G_1}(x_1, x_3 | x_2, x_4, R = 0101) = p(x_1 | x_2, x_4, R = 0101)\, p(x_3 | x_2, x_4, R = 0101),
\]
while under $G_2$, MMG implies a different imputation model
\[
p_{G_2}(x_1, x_3 | x_2, x_4, R = 0101) = p(x_1 | x_2, R = 0101)\, p(x_3 | x_2, x_4, R = 0101).
\]
The two imputation models do not agree with each other when $\bar{r}$ does not form connected components, even though the MMGs have the same neighborhood for $\bar{r}$.
\end{example}


\section{Simulation Study}  
\label{sec::simulation}


In this section we perform simulation studies to investigate the performance of the MMG. We first generate data from a Gaussian graphical model and examine the performance of G-MMG. Next, we generate data from a mixture of Gaussian graphical models and apply MP-MMG. In both scenarios, missing values are introduced under MCAR and MAR mechanisms. Note that MMG identifies the true model under MCAR (see Theorem~\ref{thm: recovery underMCAR}) but it is not guaranteed to find the correct model under MAR; the experiments on MAR are intended to investigate the robustness of MMG.

For each scenario, we generate 100 samples of size $n = 2000$. Our goal is to estimate the marginal medians of the variables. We compare four imputation strategies, including complete-case (CC) analysis, MICE \citep{vanbuuren}, missForest \citep{stekhoven2012missforest}, and the proposed G-MMG/MP-MMG. For MICE and G-MMG, we perform multiple imputation with $m = 20$ completed data sets. Simulation details are in Appendix \ref{appendix: simdetails}.

\subsection{G-MMG}

\begin{figure}
    \centering
    \includegraphics[width=\linewidth]{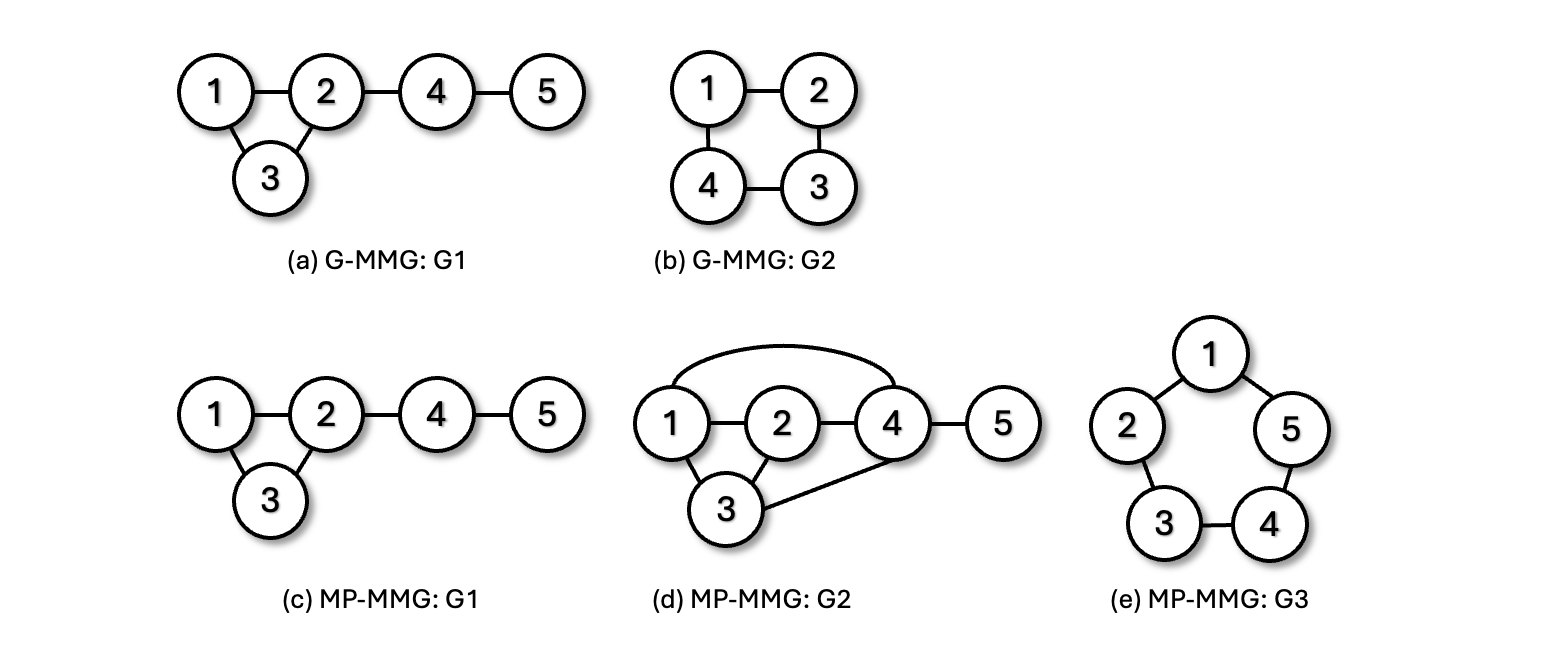}
    \caption{Graph structures in the simulation study. }
    \label{fig:simfig}
\end{figure}

We first consider data generated from Gaussian graphical models using graphs from panels (a) and (b) in Figure~\ref{fig:simfig}, with missing values introduced under MCAR and MAR. Figure~\ref{fig:simcomb} shows the median estimates across methods. The empirical results are consistent with Theorem \ref{thm: recovery underMCAR}. When the missingness mechanism is MCAR, all methods achieve comparable median estimates. However, CC analysis exhibits the highest variability among all methods, highlighting its reduced efficiency due to discarding incomplete observations.

Under MAR, the performance gaps between methods become more apparent. CC continues to show the greatest variability, while MICE often produces biased estimates. MissForest also yields estimates with both higher bias and greater variability relative to G-MMG. In comparison, G-MMG maintains both low bias and low variability across graph structures. This demonstrates the advantages of leveraging graphical structure in the imputation: G-MMG recovers efficiency lost in CC while simultaneously mitigating the biases present in MICE and missForest.

\subsection{MP-MMG}
\begin{figure}
    \centering
    \includegraphics[width=\linewidth]{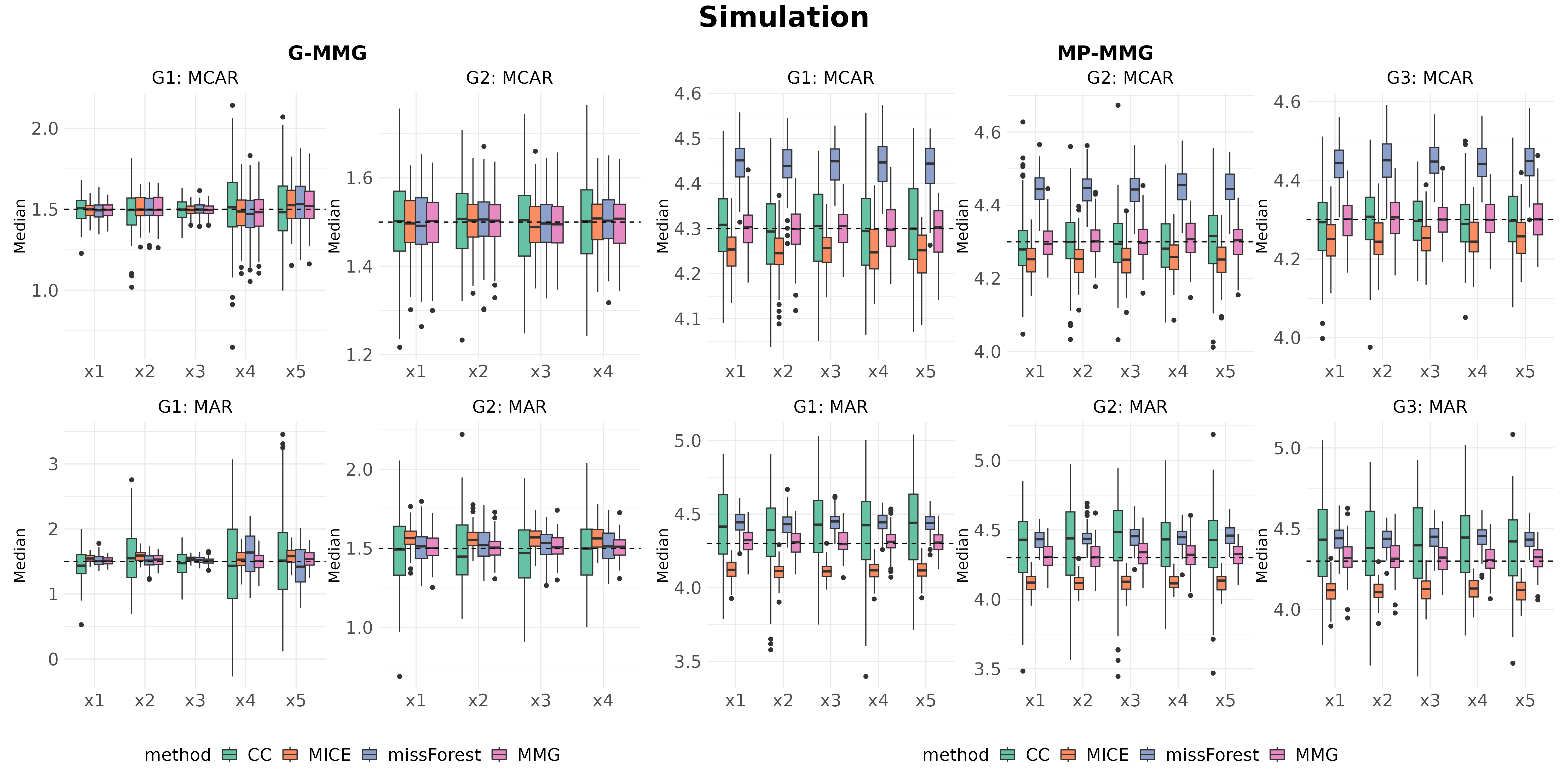}
    \caption{Median estimates across imputation methods (MMG, MICE, missForest, and CC analysis) under MCAR and MAR mechanisms. Boxplots are calculated from 100 simulation trials at sample sizes n = 2000. Left: Results from the first simulation setting where data are generated from the Gaussian graphical model. The horizontal dashed line represents the true median value of 1.5. Right: Results from the first simulation setting where data are generated from the a mixture of Gaussian graphical models. The horizontal dashed line marks the empirical median value.}
    \label{fig:simcomb}
\end{figure}

The right panels of Figure~\ref{fig:simcomb} show results for data generated from a mixture of two Gaussian graphical models, evaluated using three different working graphs in panels (c)–(e) of Figure~\ref{fig:simfig}. Despite the fact that the data-generating process is not itself a graphical model, MP-MMG produces robust estimates under both MCAR and MAR. Across all graphs, MP-MMG achieves low bias and variability, with boxplots concentrated around the true medians.

By contrast, other methods exhibit unstable performance in terms of bias and variability, with systematic deviations from the empirical median becoming evident in more complex mixture scenarios. The relative performance differences are particularly striking when the data-generating mechanism departs from simple Gaussian graphical models: CC and MICE suffer from systematic bias, while missForest shows high variability. MP-MMG, however, adapts effectively to the underlying heterogeneity and consistently produces estimates aligned with the empirical medians.

The key reason why MMG methods perform well in these settings is that their imputation submodels are better aligned with the structural features of the data. When the underlying model is simple (e.g., the left two columns of Figure~\ref{fig:simcomb}), all methods perform reasonably well. However, when the model is more complex (e.g., the right three columns of Figure~\ref{fig:simcomb}), MMG demonstrates clear advantages in both bias reduction and efficiency, underscoring its robustness to more challenging data-generating processes.




\section{Application to NACC data}   
\label{sec::nacc}
The National Alzheimer’s Coordinating Center (NACC) Uniform Data Set (UDS) is a comprehensive resource for Alzheimer’s research collected since 2005. This longitudinal data set contains subject demographics, clinical dementia rating (CDR), and a battery of neuropsychological tests across different cognitive domains. We use UDS Version 3.0, which includes data collected between 2005 and 2024. The data can be requested from the National Alzheimer’s Coordinating Center website (\url{https://naccdata.org}).

\subsection{Data}
Our primary objective is to impute missing values and model dementia progression using longitudinal neuropsychological test scores alongside demographic variables. Following the recommendations of \citet{weintraub2018version}, we select eight neuropsychological tests covering five cognitive domains: episodic memory, attention, language, processing speed, and executive function. A detailed description of the five demographic covariates and eight test scores is provided in Table \ref{var:description} in Appendix~\ref{appendix: application details}.

To evaluate post-imputation validity, we incorporate the global CDR score as an external outcome. The CDR is a clinician-rated measure of cognitive and functional impairment, widely used to stage dementia severity into five categories: 0 (normal cognition), 0.5 (mild cognitive impairment/MCI), 1 (mild impairment), 2 (moderate impairment), and 3 (severe impairment). Since the CDR score is fully observed, it provides a natural benchmark for assessing dementia progression.

To simplify the analysis, we restrict attention to individuals with records from both their baseline visit and the subsequent year. The resulting data set contains $n = 27{,}070$ individuals with 351 distinct missing data patterns, of which only 5,739 are fully observed. Thus, complete-case analysis would discard nearly 80\% of the available data. Importantly, the missingness in NACC is not random. For example, codes 96' or 996' indicate failure to complete a test due to cognitive or behavioral difficulties, affecting measures such as the Trail Making Test A/B, Multilingual Naming Test, Number Span Test, and Craft Story 21 Recall. This suggests that missingness is often associated with cognitive impairment, violating the MCAR assumption. Additionally, the 2015 transition to a new neuropsychological test battery introduced systematic instrument replacement, creating further sources of missingness.

To address these challenges, we apply the proposed MMG approach using the Mixture of Binomial Product Experts (MBPE) model \citep{suen2023modeling}, which is well suited for multivariate bounded discrete data. Implementation is based on functions adapted from the \href{https://github.com/danielsuen/mixturebpe}{\texttt{mixturebpe}}
 package.

\subsection{Model}
\begin{figure}[t]
    \centering
    \includegraphics[width=\linewidth]{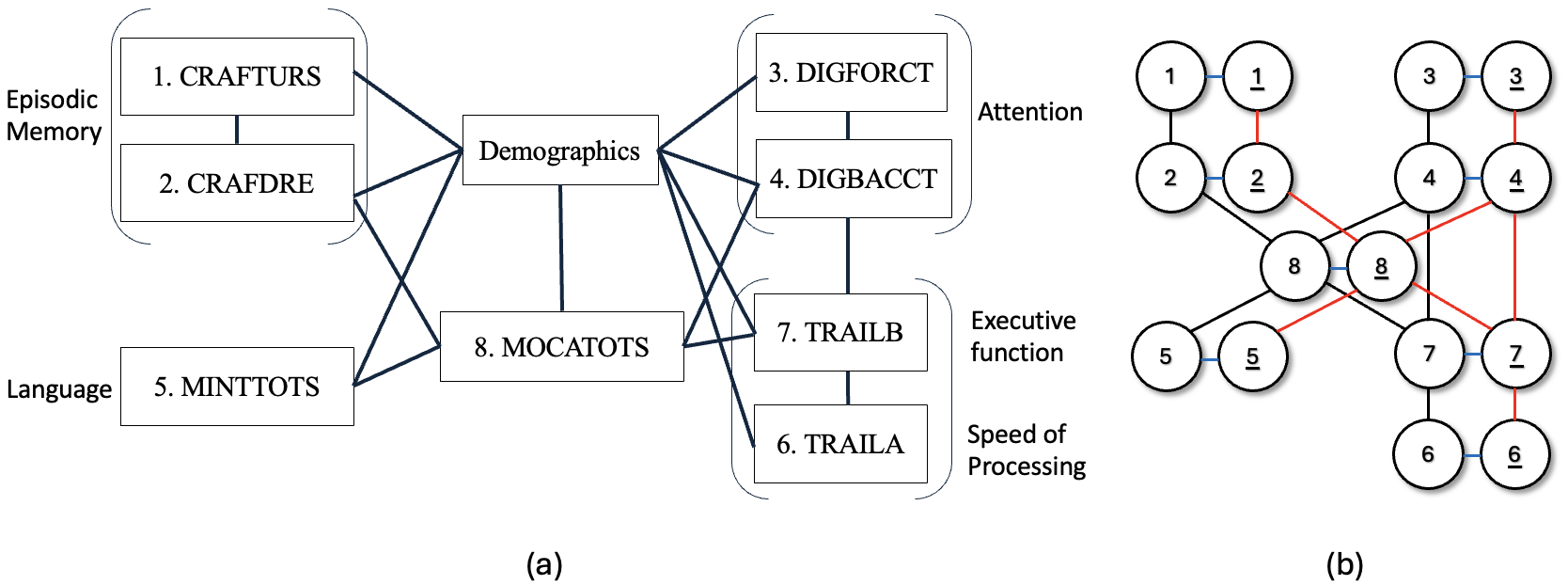}
    \caption{(a) Estimated graph using graphical lasso (glasso) based on first-year test scores. Demographic variables are assumed to connect to each test. Each test score is grouped by cognitive domain: episodic memory (CRAFTURS, CRAFTDRE), attention (DIGFORCT, DIGBACCT), language (MINTTOTS), executive function (TRAILB), and speed of processing (TRAILA). MOCATOTS is an overall cognitive assessment. (b) Extension of the dependency structure among test scores to two-year data. Nodes 1–8 represent the same first-year test scores, and nodes with underscores  denote corresponding second-year test scores. Blue edges connect each first-year test to its second-year counterpart, while red edges represent associations among second-year tests. Demographic variables are excluded from the visualization for simplicity.}
    \label{fig:glasso}
\end{figure}

To obtain a suitable graph for MMG, we use the estimated dependency structure among test scores as described in Section \ref{sec::LG}. We fit the graphical lasso (glasso) on complete-case first-year test data, selecting the tuning parameter via the extended Bayesian information criterion (EBIC; \citealp{foygel2010extended}). Demographic covariates are assumed to influence all test scores and are therefore excluded from graph estimation for clarity. The resulting graphs, shown in Figure \ref{fig:glasso}, reveal the conditional dependence relationships among the neuropsychological measures and serve as the basis for constructing MP-MMG.

From this procedure, MMG reduces the original 351 response patterns to 84 representative model patterns, reflecting shared dependency structures across missingness types. We then fit MMG with the MBPE submodels, conditioning test scores on always-observed demographic covariates, with mixture weights allowed to vary by covariate profile. Parameters for each model pattern are estimated via the EM algorithm, and $L=20$ multiply imputed data sets are generated.
For benchmarking, we also apply three commonly used imputation strategies: complete-case analysis, MICE \citep{vanbuuren}, and missForest \citep{stekhoven2012missforest}.

\subsection{Results}
\begin{figure}
    \centering
    \includegraphics[width=0.95\linewidth]{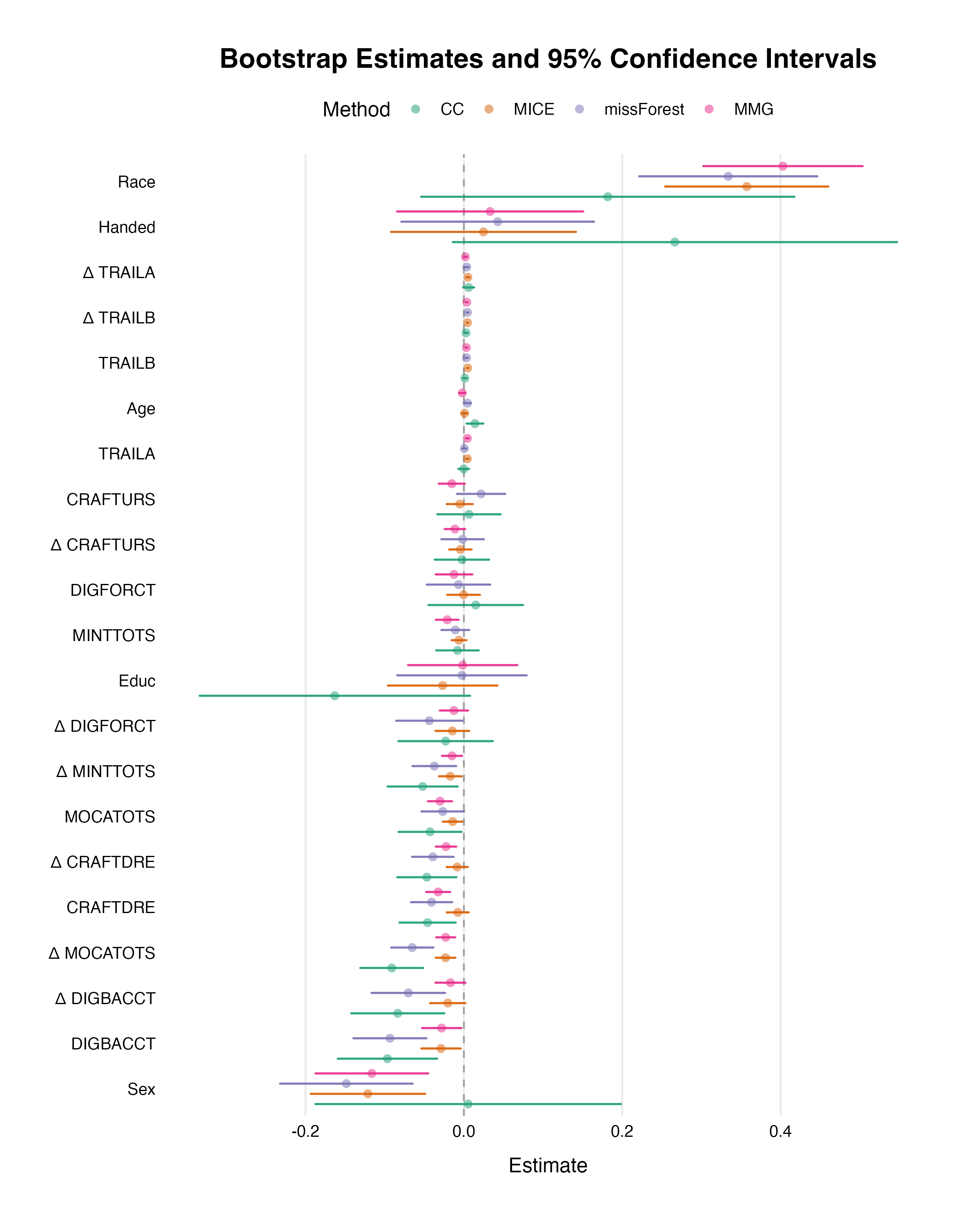}
    \caption{Coefficient estimates and 95\% confidence intervals from logistic regression under MMG, MICE, missForest, and complete-case (CC) analysis, computed from 500 bootstrap replicates. The $\Delta$ notation indicates the change between baseline and follow-up.}
    \label{fig:nacc_bootstrap}
\end{figure}

To evaluate dementia progression, we fit logistic regression models on the imputed data sets. The binary outcome is whether the CDR score increased from year 1 to year 2, with predictors including demographic variables, baseline test scores, and changes in test scores across the two years. We generate 500 bootstrap samples of the NACC data, apply each imputation method to each sample, and fit the logistic regression model on the resulting completed data sets to assess the uncertainty.

Figure \ref{fig:nacc_bootstrap} presents the resulting coefficient estimates and 95\% confidence intervals (CIs). Compared with CC analysis, MMG consistently yields narrower CIs, indicating improved efficiency. In some cases (e.g., Sex and DIGFORCT), CC estimates even diverge in sign from those obtained with MMG, MICE, and missForest, highlighting how reliance on complete cases can lead to misleading or unstable conclusions.

MMG, MICE, and missForest generally agree in coefficient signs, but missForest often exhibits wider intervals, suggesting lower efficiency. Several cognitive tests emerge as significant predictors of dementia progression. Lower baseline performance on MOCATOTS, MINTTOTS, and CRAFTDRE—assessing global cognition, language, and memory, respectively—are associated with greater risk of progression. Higher baseline scores on TRAILA and TRAILB, reflecting slower processing and weaker executive function, are also linked to increased risk. Moreover, worsening performance over time on TRAILA and TRAILB is positively associated with progression, underscoring the role of declining executive function and processing speed in disease advancement. In contrast, demographic predictors such as age and education appear to contribute little once cognitive measures are included, suggesting their effects may be mediated through test performance.

\begin{figure}[t]
    \centering
    \includegraphics[width=\linewidth]{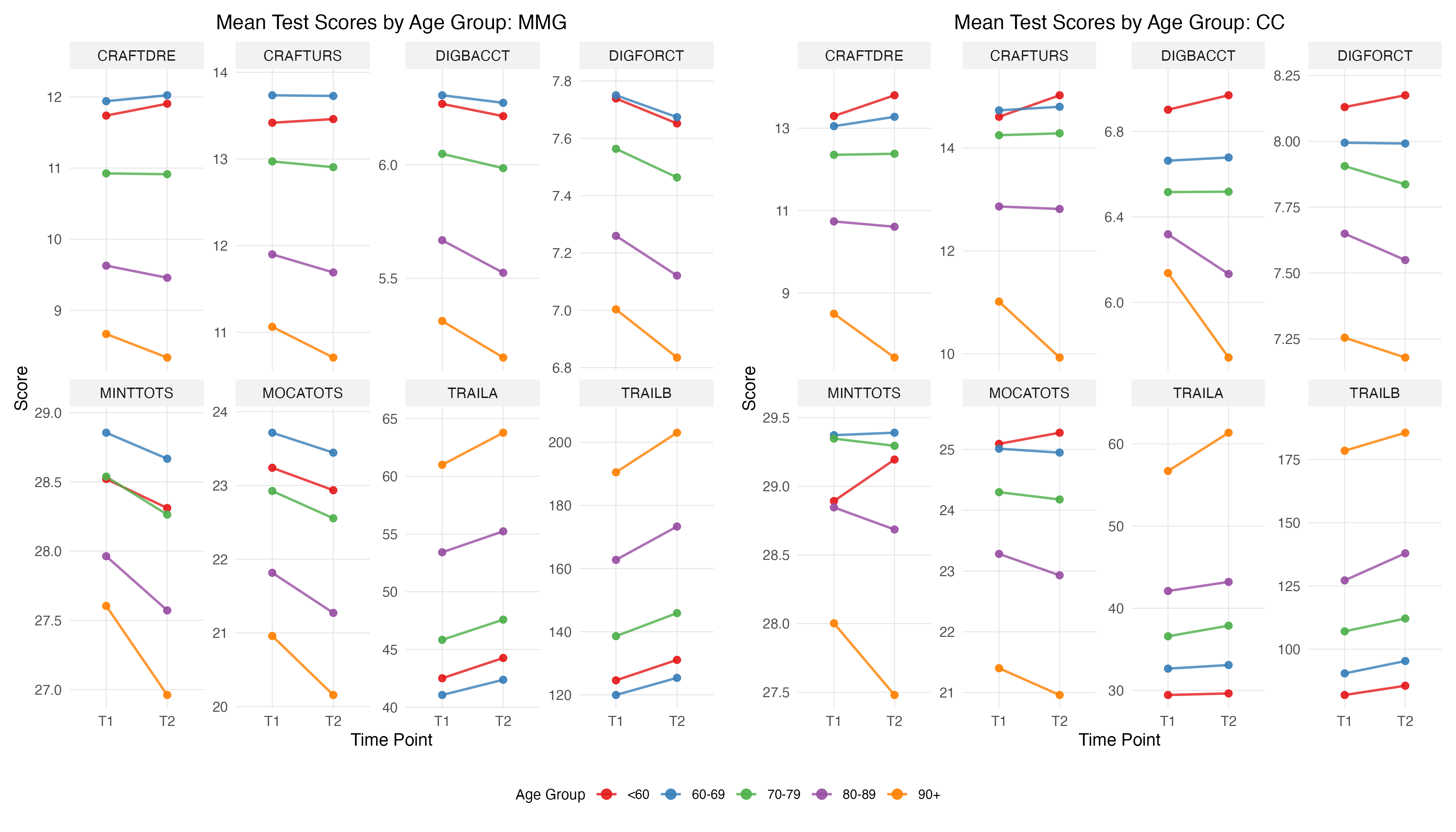}
    \caption{Mean cognitive test scores over time by age group: comparison of MMG (left) and complete-case analysis (right).}
    \label{fig:nacc_agegroup}
\end{figure}

To further investigate age-related cognitive decline, we stratify participants into five age groups: younger than 60 years, 60-69 years, 70-79 years, 80-89 years, and 90 years or older.
Figure \ref{fig:nacc_agegroup} compares age-specific trajectories of mean cognitive test scores between MMG and CC analysis. Under MMG, test scores exhibit consistent and gradual declines with increasing age, aligning with clinical expectations of cognitive aging. By contrast, CC-based trajectories display irregular and sometimes counterintuitive patterns, including apparent improvements in certain domains (e.g., MINTTOTS, MOCATOTS, DIGBACCT, DIGFORCT), {which are likely artifacts of selective dropout}.

Taken together, these results demonstrate that the choice of imputation strategy has a substantial impact on downstream analyses. MMG provides both stable effect estimation and interpretable age-related trends, whereas CC analysis risks exaggerating decline or introducing spurious patterns.

\subsection{Sensitivity Analysis}   \label{sec::sensitivity}

To assess the robustness of MMG to graph specification, we conduct a sensitivity analysis by varying the threshold parameter in the graphical lasso. This parameter controls the sparsity of the estimated dependency graph, which in turn guides the MMG imputation procedure.

\begin{figure}
\centering
\includegraphics[width=\linewidth]{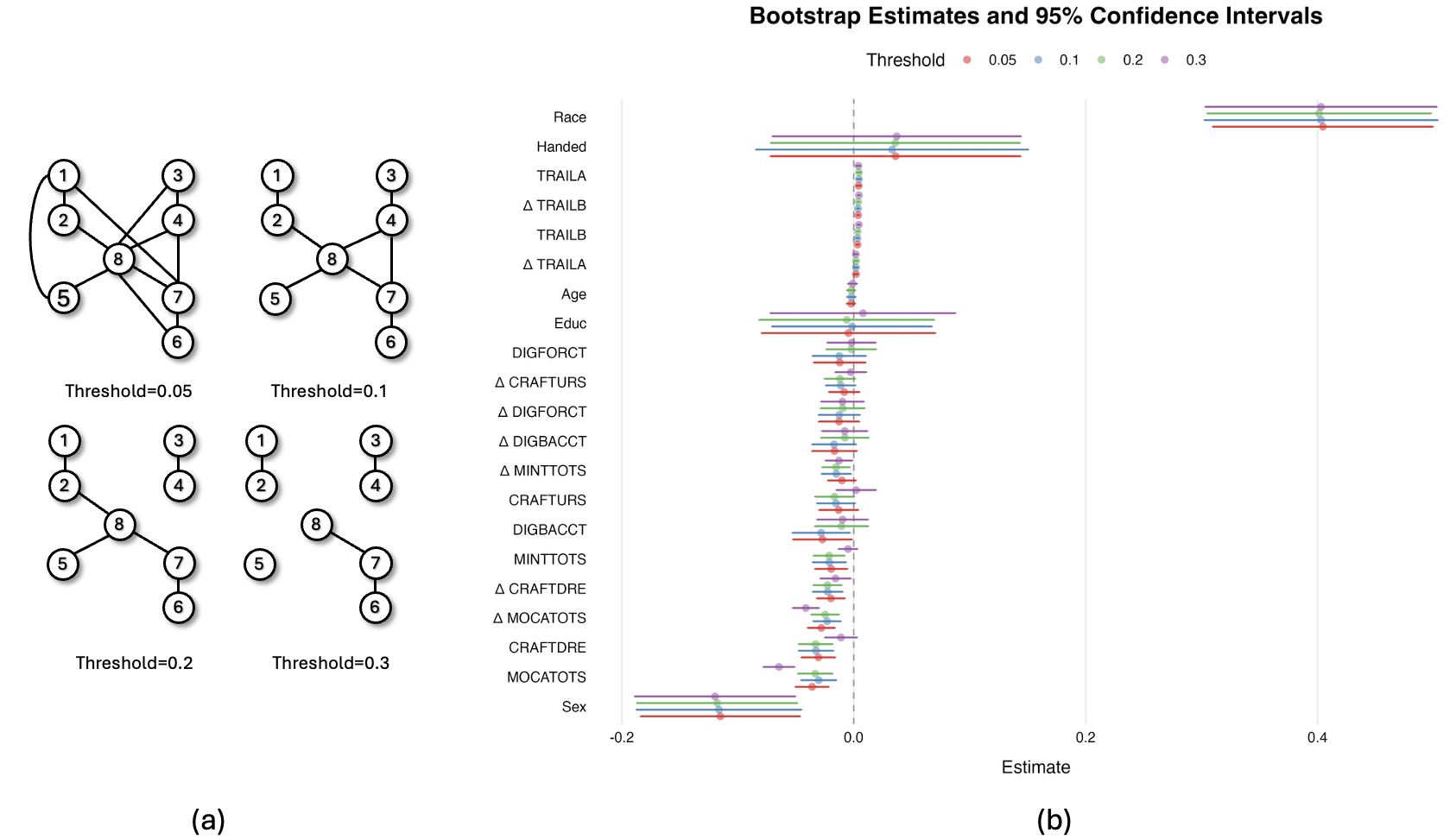}
\caption{Sensitivity analysis using the NACC data. (a) Estimated graphs based on first-year test scores under glasso thresholds of 0.05, 0.1, 0.2, and 0.3. Demographic variables are omitted for clarity. (b) Logistic regression coefficient estimates and 95\% confidence intervals across thresholds, computed from 500 bootstrap replicates.}
\label{fig:sensitivity}
\end{figure}

Panel (a) of Figure~\ref{fig:sensitivity} illustrates how the estimated graph evolves as the threshold increases. As expected, larger thresholds produce sparser graphs, with several connections disappearing at higher values. Panel (b) shows the resulting logistic regression coefficient estimates and 95\% confidence intervals. Across thresholds 0.05 to 0.2, the estimates remain highly stable, with negligible shifts in magnitude or statistical significance. At the highest threshold (0.3), however, several coefficients deviate more noticeably, reflecting the loss of important dependencies when the graph becomes overly sparse.

Overall, these findings suggest that MMG delivers robust inference across a wide range of reasonable threshold values. Importantly, this stability reinforces our main conclusion: MMG improves efficiency and interpretability over complete-case analysis and alternative imputation methods, while remaining relatively insensitive to the specific choice of graph.

\section{Discussion}
In this paper, we introduce MMG, a new framework for constructing imputation models using undirected graphs. MMG leverages graphical structure to decompose the imputation task into a collection of localized submodels, enabling both flexibility and interpretability. We establish nonparametric identification under MMG and PAI, and further propose IRM, a statistical learning framework that casts imputation as an empirical risk minimization problem. IRM accommodates a wide range of modeling choices—including Gaussian, Ising, and mixture of product models—and provides a principled basis for learning in high-dimensional settings.

On the theoretical side, we show that under certain conditions CCMV reduces to PAI within the MMG framework, and that MMG consistently recovers the true model under MCAR and graph faithfulness. We also demonstrate that MMG can be used to construct regression adjustment and IPW estimators, and we develop efficiency results that link the performance of MMG to structural properties of the underlying graph.



A natural direction for future work is to explore the relationship between MMG and missing-data DAGs (m-DAGs). Although MMG and m-DAGs are derived from different graphical principles, it remains an open question whether an m-DAG can be constructed to align with a given MMG under suitable conditions. Establishing such a connection would create a bridge between the locality and modularity of MMG and the causal interpretability of m-DAGs, potentially leading to hybrid frameworks that exploit the strengths of both paradigms.

Another extension concerns the design of PAI. In this paper we adopt the inclusive form of PAI, in which all observations containing the relevant variables are used. Alternative approaches could involve restricted PAI that prioritize observations with missingness patterns similar to the target pattern, perhaps defined via similarity metrics or proximity criteria. Such restrictions could improve computational efficiency and statistical accuracy by focusing on the most relevant cases, but at the possible cost of losing efficiency. Future work should therefore aim to develop restricted PAI that retain efficiency gains while preserving the minimal information required for valid inference.

Finally, the proposed IRM framework opens the door to integration with deep generative models, such as variational autoencoders and energy-based models. While there has been growing interest in combining graphical models with neural architectures—for example, structured variational autoencoders \citep{johnson2016composing}, neural graphical models \citep{shrivastava2023neuralgraphicalmodels}, and GRAPE \citep{you2020handling}—comparatively little work has focused on architectures tailored to missing data imputation with theoretical guarantees. This gap highlights a promising research avenue: developing MMG-compatible deep models whose generative structures respect local graph neighborhoods. Such integration could yield powerful, scalable methods for handling complex missingness while maintaining the robustness and interpretability that MMG provides.


\section*{Acknowledgements}
{We thank Daniel Suen and Yikun Zhang for useful comments on this paper. YY is supported by 2141808. YC is supported by NSF grants DMS-1952781, 2112907, 2141808, and NIH U24-AG07212.
.}

\begin{appendix}

\section{Connected Pattern and Model Pattern}
\label{appendix: model pattern}
We begin with a simple example to illustrate the concepts of \emph{connected patterns} and \emph{model patterns}, and then introduce their formal definition.

Consider $d=3$ variables $X_1, X_2, X_3$ whose dependence structure is given by the undirected graph $G$ of $X_1 - X_2 - X_3$. A \emph{response pattern} $r \in \{0,1\}^3$ indicates which variables are observed ($1$) and which are missing ($0$). For instance, the pattern $r = 010$ means that $X_1$ and $X_3$ are missing while $X_2$ is observed. In the graph $G$, the set of missing variables $\{X_1, X_3\}$ is not connected because they are separated by the observed variable $X_2$.  
Thus, the missing variables split into two connected components: $\{X_1\}$ and $\{X_3\}$. We represent each component by a binary indicator vector:
\[
s_1 = 100, \quad s_2 = 001.
\]
These are the \emph{connected patterns} corresponding to $r$. For each connected pattern $s$, the Markov boundary in $G$ is
\[
N_G(s_1) = \{X_2\}, \quad N_G(s_2) = \{X_2\}.
\]
and $\bar{N}_G(s_1) = \{X_1, X_2\}$ and $\bar{N}_G(s_2) = \{X_2, X_3\}$. Under the Principle of Available Information (PAI), the imputation submodel for $s$,
\[
p( x_s | x_{N_G(s)}, R = r ),
\]
is identifiable from all observations where the variables in $\bar{N}_G(s)$ are observed.
Consequently, many different response patterns may share the same $\overline{N}_G(s)$ and hence correspond to the same \emph{model pattern}. For instance, in the given graph $G$, the response patterns $r\in\{111, 101, 100, 001\}$ all correspond to the model pattern $111$. It is also possible for a single response pattern to correspond to multiple model patterns. For instance, with response pattern $r = 010$, the missing variables $X_1$ and $X_3$ are disconnected, which forms two distinct connected patterns and yield two model patterns: $110$ and $011$. We use the term `model pattern' because these patterns determine the set of observations used for estimation. For instance, the model pattern $011$ corresponds to fitting the imputation model using all cases where $R \geq 011$. Here are the formal definitions of connected pattern and model pattern. 

\begin{definition}[Connected Pattern]
Let $G = (V,E)$ be an undirected graph for variables $X_1, \dots, X_d$.  
Given a response pattern $r \in \{0,1\}^d$, let $\bar{r} = \mathbf{1}_d - r$ denote the missingness vector.  
A \emph{connected pattern} is a binary indicator vector $s \in \{0,1\}^d$ corresponding to a single connected component of the subgraph of $G$ induced by $\{ j : \bar{r}_j = 1 \}$, with $s_j = 1$ if and only if $j$ is in that component and $s_j = 0$ otherwise.
\end{definition}


\begin{definition}[Model Pattern]
Let $s$ be a connected pattern. The \emph{model pattern} associated with $s$ is the binary vector $m \in \{0,1\}^d$ such that
\[
m_j =
\begin{cases}
1, & j \in \overline{N}_G(s), \\
\texttt{NA}, & \text{otherwise},
\end{cases}
\]
where $m_j = 1$ indicates variables that must be observed for estimating the imputation submodel of $s$, and $\texttt{NA}$ denotes variables whose observed/missing status is irrelevant for this submodel.
\end{definition}


The connected pattern provides an easy way to define model's parameter. 
The model pattern allows us to group different response patterns 
into the same model pattern and ease the computation. 
This grouping reflects PAI: only the variables in $\bar{N}_G(s)$ need to be observed for estimating the imputation submodel associated with $s$.

\section{Simulation Details}
\label{appendix: simdetails}
\subsection{G-MMG}
Independent and identically distributed data $X$ are generated from an undirected Gaussian graphical model $X\sim N_p(\mu, \Sigma)$, where $\mu = 1.5 \cdot 1_p$ is a $p$-dimensional mean vector with all entries equal to 1.5, and $\Sigma$ is the covariance matrix.  The precision matrix $K =\Sigma^{-1} \in \mathbb{R}^{p \times p}$ encodes the conditional independence structure of the graphs in Figure~\ref{fig:simfig}, where $K_{uv} = 0$ indicates that there is no edge between vertices $u$ and $v$. The precision matrices for the two graphs are 
\[
\begin{array}{cc}
\Omega_1 =
\begin{bmatrix}
1 & 0.6 & 0.3 & 0   & 0    \\
0.6 & 1 & 0.4 & 0.3 & 0    \\
0.3 & 0.4 & 1 & 0   & 0    \\
0 & 0.3 & 0 & 1 & 0.9    \\
0 & 0   & 0 & 0.9 & 1     
\end{bmatrix}
&
\Omega_2 =
\begin{bmatrix}
1 & 0.6 & 0   & 0.4 \\
0.6 & 1 & 0.3 & 0  \\
0 & 0.3 & 1 & 0.6 \\
0.4 & 0 & 0.6 & 1
\end{bmatrix}
\end{array}
\]
To simulate missing data under MCAR, each entry in $X$ is independently masked with probability $\rho = 0.2$. To prevent unit nonresponse (\ie, rows with all entries missing), the masking process is repeated until every observation contains at least one observed value. To introduce missing values under MAR, we consider all possible $2^p - 1$ response patterns except the unit nonresponse pattern. For each pattern $r$, we specify a logistic model for the probability that an individual is assigned to pattern $r$, conditional on the observed components in that pattern, \ie $p(R=r|X) \propto  \text{expit}(\beta_0+\beta_r^TX_r)$, where $\beta_0=-1$ is the intercept, and $\beta_r=1_r$ is a vector of ones with length equal to the number of observed variables in the pattern $r$. The probabilities across all patterns are normalized to sum to one. A response pattern is then sampled independently for each observation based on these probabilities. We set the corresponding entries in $X$ to missing according to the zero entries in each pattern.

\subsection{MP-MMG}
We generate data from a mixture of two Gaussian graphical models, with mixture proportions $0.34$ and $0.66$. Each component corresponds to a multivariate normal distribution with distinct mean vectors and identity precision matrices. Specifically, for each observation $X_i \in \mathbb{R}^p$, a latent class label $Z_i \in \{1, 2\}$ is first sampled with $p(Z_i = 1) = 0.34$ and $p(Z_i = 2) = 0.66$. Conditional on $Z_i = k$, the observation is drawn from $N_p(\mu^{(k)}, I_p)$, where $\mu^{(1)} =  (0, \dots, 0)^\top$ and $\mu^{(2)} =  (5, \dots, 5)^\top$. Given the data generation process, we employed the norm method in MICE which is a parametric imputation technique,  because its underlying assumption of multivariate normality is consistent with the true nature of each component in our mixture model. We also implemented the nonparametric missForest method for comparison. In what follows, we detail how to implement the MP-MMG. First, we estimate the mixture parameters using the EM algorithm, and second, we perform the imputation using those estimated parameters. Algorithm \ref{alg:mpmmg-em} introduces the EM algorithm for parameter estimation, and Algorithm \ref{alg:mpmmg-imp} details the imputation procedure.

In the MP-MMG simulation setting, $\phi$ is taken to be the univariate Gaussian density. Accordingly, the M-step updates parameters $\theta_{k,j}=(\mu_{k,j}, \sigma_{k,j}^2)$ with closed forms:
\[
        \mu^{(t+1)}_{k,j} = \frac{\sum_{i=1}^{n_m} \gamma^{(t)}_{i,k} X_{i,j}}{\sum_{i=1}^{n_m} \gamma^{(t)}_{i,k}}, 
        \quad
        \sigma_{k,j}^{2\,(t+1)} = \frac{\sum_{i=1}^{n_m} \gamma^{(t)}_{i,k} \left(X_{i,j} - \mu_{k,j}^{(t+1)}\right)^2}{\sum_{i=1}^{n_m} \gamma^{(t)}_{i,k}}.
\]
In Step 7 of Algorithm~\ref{alg:mpmmg-imp}, missing variables are imputed by sampling from $N(\mu_{k^*,j}, \sigma_{k^*,j}^2)$.

\begin{algorithm}[h]
\caption{MP-MMG: EM Algorithm}
\label{alg:mpmmg-em}
\begin{algorithmic}[1]
\Require Data matrix $X$, response patterns $R$, graph $G$, set of model patterns $\mathcal{M}$, number of mixture components $K$
\For{each $m \in \mathcal{M}$}
    \State Identify $\bar{N}_G(m) = m \cup N_G(m)$ and extract submatrix $X^{(m)} = \{ X_i : R_i \ge \bar{N}_G(m) \}$
    \State Initialize mixture weights $\pi^{(0)}_k$ and parameters $\theta_{k,j}^{(0)}$ for all $j \in \bar{N}_G(m), k=1,\dots,K$
    \State Set iteration counter $t \gets 0$
    \Repeat
        \State \textbf{E-step:} For $i=1,\dots, n_m$ and $k = 1,\dots,K$, compute
        \[
        \gamma^{(t)}_{i,k} \;=\; \frac{\pi^{(t)}_k \prod_{j \in \bar{N}_G(m)} \phi(X_{i,j}; \theta_{k,j}^{(t)})}
        {\sum\limits_{k'=1}^K \pi^{(t)}_{k'} \prod_{j \in \bar{N}_G(m)} \phi(X_{i,j}; \theta_{k',j}^{(t)})}
        \]
        \State \textbf{M-step:} Update \[\pi^{(t+1)}_k = \frac{1}{n_m} \sum_{i=1}^{n_m} \gamma^{(t)}_{i,k},\] and for each 
        $j \in \bar{N}_G(m)$, update $\theta_{k,j}^{(t+1)}$ by maximizing 
        $\sum_{i=1}^{n_m} \gamma_{i,k}^{(t)}\log\phi(X_{i,j};\theta_{k,j})$.
        \State $t \gets t + 1$
    \Until{Convergence}
    \State \Return $\{\pi_k, \theta_{k,j}\}$ for each $m$
\EndFor
\end{algorithmic}
\end{algorithm}

\begin{algorithm}[h]
\caption{MP-MMG: Imputation}
\label{alg:mpmmg-imp}
\begin{algorithmic}[1]
\Require Data matrix $X$, response patterns $R$, graph $G$, estimated parameters $\{\pi_k, \theta_{k,j}\}$ for each $m\in\mathcal{M}$
\For{each incomplete observation $i$}
    \State Determine the set of model pattern(s) $\mathcal{M}_i$ and retrieve fitted parameters
    \For{each $m \in \mathcal{M}_i$}
    \State Identify the observed variables $O_m$ and missing variables $B_m$
    \State Compute $w_k = \pi_k \prod_{j \in O_m} \phi(X_{i,j}; \theta_{k,j})$ and normalize $w_k \leftarrow w_k / \sum_{\ell=1}^K w_\ell$
    \State Sample $k^*$ based on the normalized weights $\{w_1,\dots,w_K\}$
    \State For each $j\in B_m$, impute $X_{i,j} \sim \phi(\cdot; \theta_{k^*, j})$
    \EndFor
\EndFor
\end{algorithmic}
\end{algorithm}

\section{Variable Descriptions in the NACC Data Set}
\label{appendix: application details}
\begin{table}[H]
    \begin{adjustbox}{width=1.15\textwidth,center}
    \begin{tabular}{lcl}
    \toprule
    \text{Variable} & Domain & Description \\
    \toprule
    Sex & Demographics  & 1=Female, 0 = Male\\
    Race & Demographics  & 1 = White, 0 = Others\\
    Education & Demographics  & 1 = Bachelor's degree or more, 0 = less than Bachelor\\
    Handedness & Demographics  & 1 = Right-handed, 0 = Others\\
    Age & Demographics & Subject's age at visit\\
    CRAFTURS & Memory  & Craft Story 21 recall immediate paraphrase, total units  (0-25) \\
    CRAFTDRE & Memory  & Craft Story 21 recall delayed paraphrase, total units (0-25) \\
    DIGFORCT & Attention  & Number span test forward, total correct trials  (0-25)\\
    DIGBACCT & Attention  & Number span test backward, total correct trials (0-25)\\
    MINTTOTS & Language  & Multilingual naming test, total score (0-32)\\
    TRAILA & Speed of processing  & Trail making test part A, time (0-150)\\
    TRAILB & Executive function & Trail making test part B, time  (0-300)\\
    MOCATOTS & Overall cognitive impairment  & Montreal Cognitive Assessment, total raw score (0-30)\\
    \toprule
    \end{tabular}
    \end{adjustbox}
    \caption{Description of selected variables from the NACC data. Education, race, and handedness are recoded into binary variables. Demographic variables for analysis are fully observed while the test scores are subject to missingness. The range of test scores is shown in parentheses. }
    \label{var:description}
\end{table}

\section{Proofs}
\label{appendix: proof}

\subsection{Proof of Theorem \ref{thm: nonp-identify}}
By the pattern mixture model formulation, 
the joint PDF/PMF of $X,R$ can be decomposed into
$$
p(x,r)  = p(x_{\bar r}|x_r,r) p(x_r,r)
$$
and the only un-identifiable quantity is $p(x_{\bar r}|x_r,r)$. Thus, as long as MMG and PAI do not put any restriction on $p(x_r,r)$ and identify $p(x_{\bar r}|x_r,r)$, they together achieve nonparametric identification.
Clearly, MMG and PAI do not put any restriction on the observed data $p(x_r,r)$, so all we need is to show that $p(x_{\bar r}|x_r,r)$ can be identified from data.

By the MMG formulation in equation \eqref{eq::MMG1},
the extrapolation density is decomposed as 
$$
p(x_{\bar{r}} | x_r, R = r) = \prod_{k=1}^K p(x_{s_k} | x_{N_G(s_k)}, R = r),
$$
where $s_1,\cdots, s_K$ are connected component of the variables corresponding to $\bar r$ in the graph $G$. 
Thus, identification of $p(x_{\bar r}|x_r,r)$ reduces to identification of each $p(x_{s_k} | x_{N_G(s_k)}, R = r)$. 

By the PAI formulation in equation \eqref{eq::MMG2},
$$
p(x_{s_k}|x_{N_G(s_k)}, R = r) = p(x_{s_k}|x_{N_G(s_k)},
 R_{\bar{N}_G(s_k)} =1),
$$
and each function in the right-hand-side, $p(x_{s_k}|x_{N_G(s_k)},
 R_{\bar{N}_G(s_k)} =1)$, is clearly
identifiable from data.
Thus, $p(x_{s_k}|x_{N_G(s_k)}, R = r)$ is identifiable. 
As a result, we conclude that MMG and PAI identify the extrapolation density, which completes the proof.

\subsection{Proof of Theorem \ref{thm: PAI_CCMV_monotone}}
\begin{proof}
Because the complete cases follow a graphical model faithful to the chain $G$, we have
$(X_{l+1},\dots,X_d) \perp (X_1,\dots,X_{l-1}) \mid X_l, R=1$. 
Suppose we observe $X_1,\cdots, X_l$ ($R= r^{(l)}$)
and we want to impute $X_{l+1},\cdots, X_d$.
The imputation model by CCMV is $$p(x_{l+1},\dots,x_d|x_1,\dots,x_l,R=r^{(l)})=p(x_{l+1},\dots,x_d|x_1,\dots,x_l,R=1)=p(x_{l+1},\dots,x_d|x_l,R=1).
$$

For the MMG and PAI under the same graph, the imputation model will be
$$
p(x_{l+1},\dots,x_d|x_l,R_l=R_{l+1}=\cdots=R_d=1).
$$
In the monotone missing data, $R_d=1$ means that the last variable is observed, which only occurs when all variables are observed ($R=1$). 
Thus, 
\begin{align*}
p(x_{l+1},\dots,x_d|x_l,R_l=R_{l+1}=\cdots=R_d=1)
&= p(x_{l+1},\dots,x_d|x_l,R=1),
\end{align*}
which is exactly the same as the imputation model by CCMV.
\end{proof}

\subsection{Proof of Theorem \ref{thm: recovery underMCAR}}
\begin{proof}
MMG decomposes the imputation model into submodels based on connected components
\begin{equation*}
p(x_{\bar{r}} | x_r, R = r) = \prod_{k=1}^K p(x_{s_k} | x_{N_G(s_k)}, R = r).
\end{equation*}
Under MCAR, the missingness is independent of the data, so
\begin{align*}
p(x_{s_k}|x_{N_G(s_k)}, R=r)  = p(x_{s_k}|x_{N_G(s_k)})
\end{align*}
By faithfulness, the graph $G$ encodes the true conditional independence. Thus, the product of submodels $\prod_k p(x_{s_k} | x_{N_G(s_k)})$ correctly reconstructs the underlying distribution $p(x_{\bar{r}} | x_r)$. Therefore,  the imputation model recovers the true distribution
\[
p(x_{\bar{r}} | x_r, R = r)=\prod_{k=1}^K p(x_{s_k} | x_{N_G(s_k)})=p(x_{\bar{r}} | x_r).
\]

For efficiency proof, we focus on G-MMG for simplicity. Similar efficiency arguments can be extended to  I-MMG. Without loss of generality, we consider estimating $\mu_1=\mathbb{E}[X_1]$. The complete-case analysis uses all observations with $R=1$, which yields an estimator \[
\hat{\mu}_{\text{CCA}}=\frac{\frac{1}{n}\sum_{i=1}^n I(\bR_{i}=1)\bX_{i,1}}{\frac{1}{n}\sum_{i=1}^n I(\bR_{i}=1)}.
\] Asymptotically, it holds that \[
\sqrt{n}(\hat{\mu}_{\text{CCA}}-\mu_1)\Rightarrow N\left(0, \frac{\text{Var}(X_1)}{p(R=1)}\right).
\]
On the other hand, G-MMG yields an estimator 
\[
\hat{\mu}_{\text{G-MMG}}=\frac{1}{n}\Bigg(
\sum_{i=1}^n \bR_{i,1}\bX_{i,1}+\sum_{i=1}^n  (1-\bR_{i,1})\hat{m}_i(\bX_{i,N_G(s^{(i)})})\Bigg)
\]
where $s^{(i)} = \psi(\bR_i)$ is the connected component that contains $X_1$ for the $i$-th pattern $\bR_i$, and
$\hat{m}_i(\cdot)$ is the estimator of Gaussian conditional mean $\mathbb{E}[X_1|X_{N_G(s^{(i)})}, R_{\bar{N}_G(s^{(i)})}=1]$:
\[
\hat{m}_i(X_{N_G(s^{(i)})}) = \mathbb{E}[X_1|X_{N_G(s^{(i)})}, R_{\bar{N}_G(s^{(i)})}=1]+o_p(1).
\]
Under MCAR, $\mathbb{E}[X_1|X_{N_G(s^{(i)})}, R_{\bar{N}_G(s^{(i)})}=1] =\mathbb{E}[X_1|X_{N_G(s^{(i)})}]$.
We thus have 
\begin{equation*}
    \begin{aligned}
        \sqrt{n}(\hat{\mu}_{\text{G-MMG}}-\mu_1)=\frac{1}{\sqrt{n}}\Bigg(
        \sum_{i=1}^n \bR_{i,1}\bX_{i,1}+\sum_{i=1}^n  (1-\bR_{i,1})\mathbb{E}[X_1|X_{N_G(s^{(i)})}=\bX_{i,N_G(s^{(i)})}]-\mu_1
        \Bigg)+o_p(n^{-1/2})
    \end{aligned}
\end{equation*}
By the law of total variance, $\text{Var}(\mathbb{E}[X_1|X_{N_G(s^{(i)})}])\leq \text{Var}(X_1)$, which implies
\begin{equation*}
    \begin{aligned}
\text{Var}\Bigg( \bR_{i,1}\bX_{i,1}+(1-\bR_{i,1})\mathbb{E}[X_1|X_{N_G(s^{(i)})}
&=\bX_{i,N_G(s^{(i)})}]-\mu_1\Bigg)\\
&\leq \text{Var}\Bigg(\bR_{i,1}\bX_{i,1}+(1-\bR_{i,1})\bX_{i,1}\Bigg)\\
&=\text{Var}(X_1).
    \end{aligned}
\end{equation*}
Thus, the asymptotically variance of $\hat{\mu}_{\text{G-MMG}}$ satisfies
\[
\text{AVar}(\hat{\mu}_{\text{G-MMG}})\leq \text{Var}(X_1) <\frac{\text{Var}(X_1)}{p(R=1)}=\text{AVar}(\hat{\mu}_{\text{CCA}}).
\]
Therefore, the G-MMG is asymptotically more efficient than complete-case analysis.
\end{proof}

\subsection{Derivation of Equation~\eqref{eq: regression adjustment}}   \label{sec::RA}

\begin{align*}
\mu_r &= \mathbb{E}(X_1 I(R = r))\\ 
&= \int x_1 p(x_1, r) dx_1 \\
&= \int x_1 p(x_1, x_{-1}, r) dx_1 dx_{-1} \\
&= \int x_1 p(x_1, x_{N_G(s_j)}, r) dx_1 dx_{N_G(s_j)} \quad \text{(Markov property)} \\
&= \int x_1 p(x_1 | x_{N_G(s_j)}, r) dx_1 p(x_{N_G(s_j)}, r) dx_{N_G(s_j)} \\
&=\underbrace{ \int x_1 p(x_1 | x_{N_G(s_j)}, R \geq \bar{N}_G(s_j)) dx_1}_{m_{s_j}(X_{N_G(s_j)})} p(x_{N_G(s_j)}, r) dx_{N_G(s_j)} \quad \text{(PAI)} \\
&= \mathbb{E}(m_{s_j}(X_{N_G(s_j)}) I(R = r)).
\end{align*}

\subsection{Derivation of Equation~\eqref{eq: IPW}} \label{sec::IPW}

\begin{align*}
\mu_r &= \int x_1 p(x_1 | x_{N_G(s_j)}, R \ge \bar{N}_G(s_j)) dx_1 \, p(x_{N_G(s_j)}, r) dx_{N_G(s_j)} \\
&= \int x_1 p(x_1, x_{N_G(s_j)}, R \ge \bar{N}_G(s_j)) \frac{p(x_{N_G(s_j)}, R = r)}{p(x_{N_G(s_j)}, R \ge \bar{N}_G(s_j))} dx_{N_G(s_j)} dx_1 \\
&= \int x_1 \underbrace{\frac{p(R = r | x_{N_G(s_j)})}{p(R \ge \bar{N}_G(s_j) | x_{N_G(s_j)})}}_{\mathcal{O}_r(X_{N_G(s_j)})} p(x_1, x_{N_G(s_j)}, R \ge \bar{N}_G(s_j)) dx_{N_G(s_j)} dx_1 \\
&= \int x_1 \mathcal{O}_r(X_{N_G(s_j)}) p(x_1, x_{N_G(s_j)}, R \ge \bar{N}_G(s_j)) dx_{N_G(s_j)} dx_1 \\
&= \mathbb{E}[X_1 \cdot \mathcal{O}_r(X_{N_G(s_j)}) \cdot I(R \ge \bar{N}_G(s_j))].
\end{align*}

\subsection{Proof of Theorem \ref{thm: EIF}}
\begin{proof}
We rewrite $\mu_{s_j}$ as
\begin{align*}
\mu_{s_j} 
&= \mathbb{E}[X_1 \cdot \mathcal{O}_{s_j}(X_{N_G(s_j)}) \cdot I(R \ge \bar{N}_G(s_j))]\\
&= \int x_1 \cdot \mathcal{O}_{s_j}(X_{N_G(s_j)}) \cdot I(r \ge \bar{N}_G(s_j)) \cdot p_0(x_1, x_{N_G(s_j)}, r) \, d x_{N_G(s_j)} \, d x_1 \, dr
\end{align*}
where $p_0(\cdot)$ denote the true model. We consider a pathwise perturbation
\[
p_\epsilon(x_1, x_{N_G(s_j)}, r) = p_0(x_1, x_{N_G(s_j)}, r) \left[ 1 + \epsilon \cdot g(x_1, X_{N_G(s_j)}, r) \right], 
\]
where $g(\cdot)$ satisfies
\[
\int p_0(x_1, x_{N_G(s_j)}, r) \cdot g(x_1, x_{N_G(s_j)}, r) \, d x_{N_G(s_j)} \, d x_1 \, dr = 0.
\]

The efficient influence function $\mathrm{EIF}(x_1, x_{N_G(s_j)}, r)$ satisfies $\mathbb{E} \left[ \mathrm{EIF}(X_1, X_{N_G(s_j)}, R) \right] = 0$
and
\[
\lim_{\epsilon \to 0} \frac{\mu_{s_j}^\epsilon - \mu_{s_j}}{\epsilon}
= \int \mathrm{EIF}(x_1, x_{N_G(s_j)}, r) \cdot p_0(x_1, x_{N_G(s_j)}, r) \cdot g(x_1, x_{N_G(s_j)}, r) \, dx_1 \, dx_{N_G(s_j)} \, dr,
\]
where $\mu_{s_j}^\epsilon$ is under the model $p_\epsilon$. 

Under $p_\epsilon$, we also have the perturbed odds $\mathcal{O}_{s_j,\epsilon}(x_{N_G(s_j)})$. We define
\[
\Delta \mathcal{O}_{s_j}(x_{N_G(s_j)}) \equiv \mathcal{O}_{s_j,\epsilon}(x_{N_G(s_j)}) - \mathcal{O}_{s_j}(x_{N_G(s_j)})
\]
Then,
\begin{align*}
\mu_{s_j}^\epsilon 
&= \int x_1 \cdot \mathcal{O}_{s_j,\epsilon}(x_{N_G(s_j)}) \cdot I(r \ge \bar{N}_G(s_j)) \cdot p_\epsilon(x_1, x_{N_G(s_j)}, r) \, dx_{N_G(s_j)} \, dx_1 \, dr \\
&= \mu_{s_j} 
+ \epsilon \cdot 
\underbrace{\int x_1 \cdot \mathcal{O}_{s_j}(x_{N_G(s_j)}) \cdot I(r \ge \bar{N}_G(s_j)) \cdot p_0(x_1, x_{N_G(s_j)}, r) \cdot g(x_1, x_{N_G(s_j)}, r) \, dx_{N_G(s_j)} \, dx_1 \, dr}_{\text{Part A}} \\
&\quad + \underbrace{\int x_1 \cdot \Delta \mathcal{O}_{s_j}(x_{N_G(s_j)}) \cdot I(r \ge \bar{N}_G(s_j)) \cdot p_0(x_1, x_{N_G(s_j)}, r) \, dx_{N_G(s_j)} \, dx_1 \, dr}_{\text{Part B}} + \mathcal{O}(\epsilon^2)
\end{align*}
If we can write part B into $\int p_0(x_1, x_{N_G(s_j)}, r) g(x_1, x_{N_G(s_j)}, r) f(\cdot)dx_{N_G(s_j)} \, dx_1 \, dr$, then we can combine with part A and obtain the EIF. In particular,
\begin{align*}
\Delta \mathcal{O}_{s_j}(x_{N_G(s_j)}) &= \mathcal{O}_{s_j,\epsilon}(x_{N_G(s_j)}) - \mathcal{O}_{s_j}(x_{N_G(s_j)})\\
&= \frac{p_\epsilon(\psi(\bar{r}) = s_j \mid x_{N_G(s_j)})}{p_\epsilon(r \ge \bar{N}_G(s_j) \mid x_{N_G(s_j)})}
- \frac{p_0(\psi(\bar{r}) = s_j \mid x_{N_G(s_j)})}{p_0(r \ge \bar{N}_G(s_j) \mid x_{N_G(s_j)})} \\
&= \frac{p_\epsilon(\psi(\bar{r}) = s_j, x_{N_G(s_j)})}{p_\epsilon(r \ge \bar{N}_G(s_j), x_{N_G(s_j)})}
- \frac{p_0(\psi(\bar{r}) = s_j, x_{N_G(s_j)})}{p_0(r \ge \bar{N}_G(s_j), x_{N_G(s_j)})} \\
&= \frac{1}{p_0(r \ge \bar{N}_G(s_j), x_{N_G(s_j)})} 
\Bigg(\left[
p_\epsilon(\psi(\bar{r}) = s_j, x_{N_G(s_j)}) - p_0(\psi(\bar{r}) = s_j, x_{N_G(s_j)})
\right] \\
&\quad - \mathcal{O}_{s_j}(x_{N_G(s_j)}) \cdot 
\left[
p_\epsilon(r \ge \bar{N}_G(s_j), x_{N_G(s_j)}) - p_0(r \ge \bar{N}_G(s_j), x_{N_G(s_j)})
\right]\Bigg) + \mathcal{O}(\epsilon^2)
\end{align*}
The last step is based on Taylor expansion. We then define 
\begin{align*}
\Delta p(\psi(\bar{r})= s_j, x_{N_G(s_j)}) 
&\equiv p_{\epsilon}(\psi(\bar{r}) = s_j, x_{N_G(s_j)}) -p_0(\psi(\bar{r}) = s_j, x_{N_G(s_j)}) \\
&= \epsilon \int p_0(x_1, x_{N_G(s_j)}, r) \, g(x_1, x_{N_G(s_j)}, r) \, 
I(\psi(\bar{r}) = s_j) \, dx_1dr, \\
\Delta p(r \ge \bar{N}_G(s_j), x_{N_G(s_j)}) 
&=  p_{\epsilon}(r \ge \bar{N}_G(s_j), x_{N_G(s_j)}) -  p_0(r \ge \bar{N}_G(s_j), x_{N_G(s_j)}) \\
&= \epsilon \int p_0(x_1, x_{N_G(s_j)}, r) \, g(x_1, x_{N_G(s_j)}, r) \, 
I(r \ge \bar{N}_G(s_j)) \, dx_1dr.
\end{align*}
Substitute into the expression of $\Delta \mathcal{O}_{s_j}(x_{N_G(s_j)})$:
\begin{align*}
\Delta \mathcal{O}_{s_j}(x_{N_G(s_j)}) 
&= \frac{\epsilon}{p_0(r \ge \bar{N}_G(s_j), x_{N_G(s_j)})} 
\int p_0(x_1, x_{N_G(s_j)}, r) \, g(x_1, x_{N_G(s_j)}, r) \\
&\quad \cdot \left\{ I(\psi(\bar{r}) = s_j) - \mathcal{O}_{s_j}(x_{N_G(s_j)}) \cdot I(r \ge \bar{N}_G(s_j)) \right\}
\, dx_1dr +\mathcal{O}(\epsilon^2).
\end{align*}
Therefore, part B becomes
\begin{align*}
&\int x_1 \cdot \Delta \mathcal{O}_{s_j}(x_{N_G(s_j)}) \cdot I(r \ge \bar{N}_G(s_j)) \cdot p_0(x_1, x_{N_G(s_j)}, r) \, dx_{N_G(s_j)} \, dx_1 \, dr \\
&= \int \Delta \mathcal{O}_{s_j}(X_{N_G(s_j)}) \cdot p_0(x_{N_G(s_j)}, r \geq N_G(s_j)) \cdot x_1 \cdot p_0(x_1 \mid x_{N_G(s_j)}, r \geq N_G(s_j)) \, dx_1 \, dx_{N_G(s_j)} \\
&= \int \Delta \mathcal{O}_{s_j}(X_{N_G(s_j)}) \cdot p_0(x_{N_G(s_j)}, r \geq N_G(s_j)) \cdot m_{s_j}(x_{N_G(s_j)}) \, dx_{N_G(s_j)} \\
&= \epsilon \int p_0(x_1, x_{N_G(s_j)}, r) \cdot g(x_1, x_{N_G(s_j)}, r) \left\{ I(\psi(r) = s_j) - \mathcal{O}_{s_j}(X_{N_G(s_j)})I(r \geq N_G(s_j)) \right\} \\
&\quad \cdot m_{s_j}(x_{N_G(s_j)})  dx_1dx_{N_G(s_j)} dr
\end{align*}
Combining part A and part B, we get
\begin{align*}
\mu_{s_j}^\epsilon - \mu_{s_j} 
&= \epsilon \int x_1 \cdot \mathcal{O}_{s_j}(x_{N_G(s_j)}) \cdot I(r \ge \bar{N}_G(s_j)) \cdot p_0(x_1, x_{N_G(s_j)}, r) \cdot g(x_1, x_{N_G(s_j)}, r) \, dx_{N_G(s_j)} \, dx_1 \, dr \\
&+ \epsilon \int p_0(x_1, x_{N(s_j)}, r) \cdot g(x_1, x_{N_G(s_j)}, r) \left\{ I(\psi(r) = s_j) - \mathcal{O}_{s_j}(x_{N_G(s_j)}) \cdot I(r \geq N_G(s_j)) \right\}\\
&\quad \cdot m_{s_j}(x_{N_G(s_j)}) \, dx_{N_G(s_j)} dx_1 dr +\mathcal{O}(\epsilon^2)
\end{align*}
Therefore, the efficient influence function is 
\begin{equation*}
    \begin{aligned}
        \mathrm{EIF}(X_1, X_{N_G(s_j)}, R) &= 
X_1 \cdot \mathcal{O}_{s_j}(X_{N_G(s_j)}) \cdot  I(R \ge \bar{N}_G(s_j)) 
\\
& +m_{s_j}(X_{N_G(s_j)})\Bigg(I(\psi(\bar{R}) = s_j) - I(R \ge \bar{N}_G(s_j))\cdot \mathcal{O}_{s_j}(X_{N_G(s_j)}) 
\Bigg) \\
&- \mathbb{E}[X_1 I(\psi(\bar{R}) = s_j)].
    \end{aligned}
\end{equation*}
\end{proof}

\subsection{Proof of Theorem \ref{thm: multiple robustness}}    \label{sec::MR}

\begin{proof}
In this proof, we will use some notations from
empirical process theory. 
Let $\mathbb{P}_0$ be the probability measure of $(X,R)$ that generates our data and $\mathbb{P}_n$
be the corresponding empirical measure. 
For any function $g(X,R)$, we write
$$
\mathbb{P}_0(g(X,R)) = \mathbb{E}[g(X,R)] = \mathbb{P}_0 g,\qquad
\mathbb{P}_n(g(X,R)) = \frac{1}{n}\sum_{i=1}^n[g(\bX_i,\bR_i)] = \mathbb{P}_n g.
$$

Let $\Tilde{\mu}$ be the oracle estimator that replaces estimated nuisance with true functions. 
\begin{equation*}
    \begin{aligned}
        \Tilde{\mu} &\equiv \sum_{r:r_1=1}\mathbb{P}_n X_{1}I(R=r)+\sum_{j=1}^J\Tilde{\mu}_{s_j}\\
        &= \sum_{r:r_1=1}\mathbb{P}_n X_{1}I(R=r)\\
        & +\sum_{j=1}^J \mathbb{P}_n \Bigg[X_{1}O_{s_j}(X_{N_{G}(s_j)})I(R \ge \bar{N}_G(s_j)) + m_{s_j}(X_{N_G(s_j)})\Bigg(I(\psi(\bar{R}) = s_j) - I(R \ge \bar{N}_G(s_j)) \mathcal{O}_{s_j}(X_{N_G(s_j)}) \Bigg)
        \Bigg]
    \end{aligned}
\end{equation*}
It is not hard to show that $\Tilde{\mu}\overset{p}{\to} \mu$. A direct computation shows that 
\begin{equation*}
    \begin{aligned}
\hat{\mu} - \Tilde{\mu} 
&= \mathbb{P}_n
\Bigg[
\underbrace{\sum_{j=1}^J \{X_1-m_{s_j}(X_{N_{G}(s_j)})\}\{\hat{\mathcal{O}}_{s_j}(X_{N_{G}(s_j)})- \mathcal{O}_{s_j}(X_{N_{G}(s_j)})\} I(R \ge \bar{N}_G(s_j))}_{\text{Term I}} \\
&\qquad + \underbrace{\sum_{j=1}^J \{\hat{m}_{s_j}(X_{N_{G}(s_j)})- m_{s_j}(X_{N_{G}(s_j)})\}\{I(\psi(\bar{R}) = s_j)-\mathcal{O}_{s_j}(X_{N_{G}(s_j)})\cdot I(R \ge \bar{N}_G(s_j)\}}_{\text{Term II}}\\
&\qquad -  \underbrace{\sum_{j=1}^J \{\hat{m}_{s_j}(X_{N_{G}(s_j)})- m_{s_j}(X_{N_{G}(s_j)})\}\{\hat{\mathcal{O}}_{s_j}(X_{N_{G}(s_j)})- \mathcal{O}_{s_j}(X_{N_{G}(s_j)})\}I(R \ge \bar{N}_G(s_j))}_{\text{Term III}}
\Bigg]
    \end{aligned}
\end{equation*}
We first establish the multiple robustness when the odds are correctly specified and the regression functions are misspecified, \ie, $\mathcal{O}_{s_j}=\mathcal{O}_{s_j}^*$ and  $m_{s_j}\neq m_{s_j}^*$. For term I, we define \[
g_{j,n}(r, x_1, x_{N_{G}(s_j)}):=\{x_1-m_{s_j}(x_{N_{G}(s_j)})\}\{\hat{\mathcal{O}}_{s_j}(x_{N_{G}(s_j)})- \mathcal{O}_{s_j}(x_{N_{G}(s_j)})\} I(r \ge \bar{N}_G(s_j))
\]
and 
\[
g_{j,0}(r, x_1, x_{N_{G}(s_j)}):=\{x_1-m_{s_j}(x_{N_{G}(s_j)})\}\{\mathcal{O}^*_{s_j}(x_{N_{G}(s_j)})- \mathcal{O}_{s_j}(x_{N_{G}(s_j)})\} I(r \ge \bar{N}_G(s_j))=0
\]
Term I for index $j$ becomes $\mathbb{P}_n(g_{j,n}-g_{j,0})$. Given assumptions (A1) and (A2), we have $\norm{\hat{\mathcal{O}}_{s_j}-\mathcal{O}_{s_j}}_{L_2(P)}=o_P(1)$, and $\norm{g_{j,n}-g_{j,0}}_{L_2(P)}=o_P(1)$ due to uniformly bounded. Since $\hat{\mathcal{O}}_{s_j}$ is in a Donsker class $\mathcal{F}_j$, $g_{j,n}$ is also in a Donsker class 
\[
\mathcal{F}^*_j = \left\{
[x_1-m_{s_j}(x_{N_{G}(s_j)})][h(x_{N_{G}(s_j)})- \mathcal{O}_{s_j}(x_{N_{G}(s_j)})] I(r \ge \bar{N}_G(s_j)): h\in \mathcal{F}_j
\right\}.
\]By Lemma 19.24 of \cite{van2000asymptotic}, it holds that \[
(\mathbb{P}_n-\mathbb{P}_0)(g_{j,n}-g_{j,0})=o_P(n^{-1/2}).
\]
which implies that $\mathbb{P}_ng_{j,n}=o_P(n^{-1/2})$.

For term II, we similarly define 
\[
f_{j,n}(r,x_{N_G(s_j)}):=(\hat{m}_{s_j}-m_{s_j})[I(\psi(\bar{r})=s_j)-\mathcal{O}_{s_j}\cdot I(r\geq \bar{N}_G(s_j))]
\]
and
\[
f_{j,0}(r,x_{N_G(s_j)}):=(m^*_{s_j}-m_{s_j})[I(\psi(\bar{r})=s_j)-\mathcal{O}_{s_j}\cdot I(r\geq \bar{N}_G(s_j))]
\]
Term II for index $j$ becomes 
\begin{equation*}
    \begin{aligned}
        \text{Term II}_j&=(\mathbb{P}_n-\mathbb{P}_0)(f_{j,n}-f_{j,0})+\mathbb{P}_n(m^*_{s_j}-m_{s_j})[I(\psi(\bar{R})=s_j)-\mathcal{O}_{s_j}I(R\geq \bar{N}_G(s_j))].
    \end{aligned}
\end{equation*}
Because the second part is 
$$
\mathbb{P}_n(m^*_{s_j}-m_{s_j})[I(\psi(\bar{R})=s_j)-\mathcal{O}_{s_j}I(R\geq \bar{N}_G(s_j))] = \mathbb{P}_nf_{j,0}=o_P(1),
$$
we have 
\[
\text{Term II}_j = (\mathbb{P}_n-\mathbb{P}_0)(f_{j,n}-f_{j,0}) + o_P(1).
\]
Given the assumption (A1), we have $\norm{f_{j,n}-f_{j,0}}_{L_2(P)}=o_P(1)$ and $f_{j,n}$ is in a Donsker class due to a similar reasoning as before. By Lemma 19.24 of \cite{van2000asymptotic}, it holds that \[(\mathbb{P}_n-\mathbb{P}_0)(f_{j,n}-f_{j,0})=o_P(n^{-1/2}).\]
It implies that $\mathbb{P}_nf_{j,n}=o_P(1)+o_P(n^{-1/2})=o_P(1)$.

For term III, we define 
\[
h_{j,n}:=(m_{s_j}-\hat{m}_{s_j})(\mathcal{O}_{s_j}-\hat{\mathcal{O}}_{s_j})I(r\geq \bar{N}_{G(s_j)})
\]
and \[
h_{j,0}:=(m_{s_j}-\hat{m}_{s_j})(\mathcal{O}_{s_j}-\mathcal{O}^*_{s_j})I(r\geq \bar{N}_{G(s_j)})=0.
\]
Then $h_{j,n}$ is in a Donsker class by a similar reasoning as before. Given the assumption (A1) and (A4), we have $\norm{h_{j,n}-h_{j,0}}_{L_2(P)}=o_P(1)$. By Lemma 19.24 of \cite{van2000asymptotic}, it holds that \[
(\mathbb{P}_n-\mathbb{P}_0)(h_{j,n}-h_{j,0})=o_P(n^{-1/2}).
\]
Then we have $\mathbb{P}_nh_{j,n}=\mathbb{P}_0h_{j,n}+o_P(n^{-1/2})=o_P(1)$. The last step is because $\mathbb{P}_0h_{j,n}\leq \norm{m_{s_j}-\hat{m}_{s_j}}_{L_2(P)}\norm{\mathcal{O}_{s_j}-\hat{\mathcal{O}}_{s_j}}_{L_2(P)}=o_P(1)$ under the assumption (A3).
Combining all the above results, we have \[\hat{\mu}=\Tilde{\mu}+o_p(1)=\mu+o_p(1).\] Similar proof applies when the regression functions are correctly specified and the odds are misspecified. Consequently, we have shown that $\hat{\mu}\overset{p}{\to} \mu$.    
\end{proof}

\subsection{Proof of Proposition \ref{prop1}}

\begin{proof}
Because the two patterns have the same local structure of a connected subset $s$
 that
 $r_1 \neq r_2$ such that $
r_{1,j} = r_{2,j} =0$ for all $j$ with $s_j=1$  and 
$r_{1,k} = r_{2,k} = 1$ for all $k \in N_G(s)$,
this implies $s$ is a connected component in both $r_1$ and $r_2$. 
Thus, PAI implies that 
    \[
    p(x_s|x_{N_G(s)},R=r_1) = p(x_s|x_{N_G(s)},R_{\bar{N}_G(s)}\geq 1) = p(x_s|x_{N_G(s)},R=r_2).
    \]
\end{proof}

\subsection{Proof of Proposition \ref{prop2}}
\begin{proof}
    By PAI, we have
\begin{align*}
    p(x_A | x_{N_G(A)}, R_A = 0, R_{N_G(A)} = 1) &= p(x_A | x_{N_G(A)}, R_{N_G(A)} \geq 1), \\
    p(x_B | x_{N_G(B)}, R_B = 0, R_{N_G(B)} = 1) &= p(x_B | x_{N_G(B)}, R_{N_G(B)} \geq 1).
\end{align*}
Since $A \cup N_G(A) = \bar{N}_G(A) = \bar{N}_G(B) = B \cup N_G(B)$, we have
\[
 \bar{N}_G(A) = A \cup N_G(A) = B \cup (A \setminus B) \cup N_G(A) = B \cup N_G(B),
\]
so
\begin{equation}
    N_G(B) = N_G(A) \cup (A \setminus B).
    \label{eq::NGA}
\end{equation}
    
With this, we have the following equalities
\begin{align*}
    p(x_B | x_{N_G(B)}, R_A = 0, R_{N_G(A)} = 1) 
    &= p(x_B | x_{N_G(A)}, x_{A \setminus B}, R_A = 0, R_{N_G(A)} = 1) \\
    &= \frac{p(x_A | x_{N_G(A)}, R_A = 0, R_{N_G(A)} = 1)}{p(x_{A \setminus B} | x_{N_G(A)}, R_A = 0, R_{N_G(A)} = 1) } \\
    &= \frac{p(x_A | x_{N_G(A)}, R_A = 0, R_{N_G(A)} = 1)}{\int p(x_A | x_{N_G(A)}, R_A = 0, R_{N_G(A)} = 1) dx_B} \\
    &= \frac{p(x_A | x_{N_G(A)}, R_{N_G(A)} \geq 1)}{\int p(x_A | x_{N_G(A)}, R_{N_G(A)} \geq 1) dx_B} \\
    &= \frac{p(x_A | x_{N_G(A)}, R_{N_G(A)} \geq 1)}{p(x_{A \setminus B} | x_{N_G(A)}, R_{N_G(A)} \geq 1) } \\
    &= p(x_B | x_{N_G(A)}, x_{A \setminus B}, R_{N_G(A)} \geq 1) \\
    &\overset{\eqref{eq::NGA}}{=} p(x_B | x_{N_G(B)}, R_{N_G(B)} \geq 1) \\
    &= p(x_B | x_{N_G(B)}, R_B = 0, R_{N_G(B)} = 1).
\end{align*}
\end{proof}

\end{appendix}

\vskip 0.2in
\bibliography{ref}

\begin{thebibliography}{42}
\providecommand{\natexlab}[1]{#1}
\providecommand{\url}[1]{\texttt{#1}}
\expandafter\ifx\csname urlstyle\endcsname\relax
  \providecommand{\doi}[1]{doi: #1}\else
  \providecommand{\doi}{doi: \begingroup \urlstyle{rm}\Url}\fi

\bibitem[Angelopoulos et~al.(2023)Angelopoulos, Bates, Fannjiang, Jordan, and Zrnic]{angelopoulos2023prediction}
Anastasios~N. Angelopoulos, Stephen Bates, Clara Fannjiang, Michael~I. Jordan, and Tijana Zrnic.
\newblock Prediction-powered inference.
\newblock \emph{Science}, 382\penalty0 (6671):\penalty0 669--674, 2023.

\bibitem[Bell et~al.(2014)Bell, Fiero, Horton, and Hsu]{bell2014handling}
Melanie~L Bell, Mallorie Fiero, Nicholas~J. Horton, and Chiu-Hsieh Hsu.
\newblock Handling missing data in {RCT}s; a review of the top medical journals.
\newblock \emph{BMC Medical Research Methodology}, 14\penalty0 (1):\penalty0 118, 2014.

\bibitem[Bhattacharya et~al.(2020)Bhattacharya, Nabi, Shpitser, and Robins]{bhattacharya2020identification}
Rohit Bhattacharya, Razieh Nabi, Ilya Shpitser, and James~M. Robins.
\newblock Identification in missing data models represented by directed acyclic graphs.
\newblock In \emph{Uncertainty in Artificial Intelligence}, pages 1149--1158. PMLR, 2020.

\bibitem[Burgette and Reiter(2010)]{burgette2010multiple}
Lane~F. Burgette and Jerome~P. Reiter.
\newblock Multiple imputation for missing data via sequential regression trees.
\newblock \emph{American Journal of Epidemiology}, 172\penalty0 (9):\penalty0 1070--1076, 2010.

\bibitem[Chen(2022)]{chen2022pattern}
Yen-Chi Chen.
\newblock Pattern graphs: a graphical approach to nonmonotone missing data.
\newblock \emph{The Annals of Statistics}, 50\penalty0 (1):\penalty0 129--146, 2022.

\bibitem[Chen and Sadinle(2019)]{chen2019nonparametric}
Yen-Chi Chen and Mauricio Sadinle.
\newblock Nonparametric pattern-mixture models for inference with missing data, 2019.
\newblock URL \url{https://arxiv.org/abs/1904.11085}.

\bibitem[Emmanuel et~al.(2021)Emmanuel, Maupong, Mpoeleng, Semong, Mphago, and Tabona]{emmanuel2021survey}
Tlamelo Emmanuel, Thabiso Maupong, Dimane Mpoeleng, Thabo Semong, Banyatsang Mphago, and Oteng Tabona.
\newblock A survey on missing data in machine learning.
\newblock \emph{Journal of Big data}, 8\penalty0 (1):\penalty0 140, 2021.

\bibitem[Foygel and Drton(2010)]{foygel2010extended}
Rina Foygel and Mathias Drton.
\newblock Extended bayesian information criteria for gaussian graphical models.
\newblock \emph{Advances in Neural Information Processing Systems}, 23, 2010.

\bibitem[Friedman et~al.(2008)Friedman, Hastie, and Tibshirani]{friedman2008sparse}
Jerome Friedman, Trevor Hastie, and Robert Tibshirani.
\newblock Sparse inverse covariance estimation with the graphical lasso.
\newblock \emph{Biostatistics}, 9\penalty0 (3):\penalty0 432--441, 2008.

\bibitem[Gondara and Wang(2018)]{gondara2018mida}
Lovedeep Gondara and Ke~Wang.
\newblock {MIDA}: Multiple imputation using denoising autoencoders.
\newblock In \emph{Pacific-Asia Conference on Knowledge Discovery and Data Mining}, pages 260--272. Springer, 2018.

\bibitem[Johnson et~al.(2016)Johnson, Duvenaud, Wiltschko, Adams, and Datta]{johnson2016composing}
Matthew~J. Johnson, David~K. Duvenaud, Alex Wiltschko, Ryan~P. Adams, and Sandeep~R. Datta.
\newblock Composing graphical models with neural networks for structured representations and fast inference.
\newblock \emph{Advances in Neural Information Processing Systems}, 29, 2016.

\bibitem[Li et~al.(2023)Li, Miao, Shpitser, and Tchetgen~Tchetgen]{li2023self}
Yilin Li, Wang Miao, Ilya Shpitser, and Eric~J. Tchetgen~Tchetgen.
\newblock A self-censoring model for multivariate nonignorable nonmonotone missing data.
\newblock \emph{Biometrics}, 79\penalty0 (4):\penalty0 3203--3214, 2023.

\bibitem[Little(1993)]{little1993pattern}
Roderick~J.A. Little.
\newblock Pattern-mixture models for multivariate incomplete data.
\newblock \emph{Journal of the American Statistical Association}, 88\penalty0 (421):\penalty0 125--134, 1993.

\bibitem[Malinsky et~al.(2022)Malinsky, Shpitser, and Tchetgen~Tchetgen]{malinsky2022semiparametric}
Daniel Malinsky, Ilya Shpitser, and Eric~J. Tchetgen~Tchetgen.
\newblock Semiparametric inference for nonmonotone missing-not-at-random data: the no self-censoring model.
\newblock \emph{Journal of the American Statistical Association}, 117\penalty0 (539):\penalty0 1415--1423, 2022.

\bibitem[Mattei and Frellsen(2019)]{pmlr-v97-mattei19a}
Pierre-Alexandre Mattei and Jes Frellsen.
\newblock {MIWAE}: Deep generative modelling and imputation of incomplete data sets.
\newblock In \emph{International Conference on Machine Learning}, volume~97, pages 4413--4423. PMLR, 2019.

\bibitem[Meinshausen and B{\"u}hlmann(2006)]{meinshausen2006high}
Nicolai Meinshausen and Peter B{\"u}hlmann.
\newblock High-dimensional graphs and variable selection with the lasso.
\newblock \emph{The Annals of Statistics}, 34\penalty0 (3):\penalty0 1436--1462, 2006.

\bibitem[Mohan and Pearl(2014)]{mohan2014testability}
Karthika Mohan and Judea Pearl.
\newblock On the testability of models with missing data.
\newblock In \emph{Artificial Intelligence and Statistics}, pages 643--650. PMLR, 2014.

\bibitem[Mohan and Pearl(2021)]{mohan2021graphical}
Karthika Mohan and Judea Pearl.
\newblock Graphical models for processing missing data.
\newblock \emph{Journal of the American Statistical Association}, 116\penalty0 (534):\penalty0 1023--1037, 2021.

\bibitem[Mohan et~al.(2013)Mohan, Pearl, and Tian]{mohan2013graphical}
Karthika Mohan, Judea Pearl, and Jin Tian.
\newblock Graphical models for inference with missing data.
\newblock \emph{Advances in Neural Information Processing Systems}, 26, 2013.

\bibitem[Molenberghs et~al.(1998)Molenberghs, Michiels, Kenward, and Diggle]{molenberghs1998monotone}
Geert Molenberghs, Bart Michiels, Michael~G. Kenward, and Peter~J. Diggle.
\newblock Monotone missing data and pattern-mixture models.
\newblock \emph{Statistica Neerlandica}, 52\penalty0 (2):\penalty0 153--161, 1998.

\bibitem[Molenberghs et~al.(2014)Molenberghs, Fitzmaurice, Kenward, Tsiatis, and Verbeke]{molenberghs2014handbook}
Geert Molenberghs, Garrett Fitzmaurice, Michael~G Kenward, Anastasios Tsiatis, and Geert Verbeke.
\newblock \emph{Handbook of Missing Data Methodology}.
\newblock CRC Press, 2014.

\bibitem[Murray and Reiter(2016)]{murray2016multiple}
Jared~S. Murray and Jerome~P. Reiter.
\newblock Multiple imputation of missing categorical and continuous values via bayesian mixture models with local dependence.
\newblock \emph{Journal of the American Statistical Association}, 111\penalty0 (516):\penalty0 1466--1479, 2016.

\bibitem[Nabi et~al.(2020)Nabi, Bhattacharya, and Shpitser]{nabi2020full}
Razieh Nabi, Rohit Bhattacharya, and Ilya Shpitser.
\newblock Full law identification in graphical models of missing data: Completeness results.
\newblock In \emph{International Conference on Machine Learning}, pages 7153--7163. PMLR, 2020.

\bibitem[Phung et~al.(2025)Phung, Reese, Shpitser, and Bhattacharya]{phung2025recursiveequationsimputationmissing}
Trung Phung, Kyle Reese, Ilya Shpitser, and Rohit Bhattacharya.
\newblock Recursive equations for imputation of missing not at random data with sparse pattern support, 2025.
\newblock URL \url{https://arxiv.org/abs/2507.16107}.

\bibitem[Raghunathan et~al.(2001)Raghunathan, Lepkowski, Van~Hoewyk, Solenberger, et~al.]{raghunathan2001multivariate}
Trivellore~E. Raghunathan, James~M. Lepkowski, John Van~Hoewyk, Peter Solenberger, et~al.
\newblock A multivariate technique for multiply imputing missing values using a sequence of regression models.
\newblock \emph{Survey Methodology}, 27:\penalty0 85--96, 2001.

\bibitem[Robins(1997)]{robins1997MNAR}
James~M. Robins.
\newblock Non-response models for the analysis of non-monotone non-ignorable missing data.
\newblock \emph{Statistics in Medicine}, 16\penalty0 (1):\penalty0 21--37, 1997.

\bibitem[Sadinle and Reiter(2017)]{sadinle2017itemwise}
Mauricio Sadinle and Jerome~P. Reiter.
\newblock Itemwise conditionally independent nonresponse modelling for incomplete multivariate data.
\newblock \emph{Biometrika}, 104\penalty0 (1):\penalty0 207--220, 2017.

\bibitem[Shpitser(2016)]{shpitser2016consistent}
Ilya Shpitser.
\newblock Consistent estimation of functions of data missing non-monotonically and not at random.
\newblock \emph{Advances in Neural Information Processing Systems}, 29, 2016.

\bibitem[Shpitser et~al.(2015)Shpitser, Mohan, and Pearl]{shpitser2015missing}
Ilya Shpitser, Karthika Mohan, and Judea Pearl.
\newblock Missing data as a causal and probabilistic problem.
\newblock In \emph{Uncertainty in Artificial Intelligence}, pages 802--811, 2015.

\bibitem[Shrivastava and Chajewska(2023)]{shrivastava2023neuralgraphicalmodels}
Harsh Shrivastava and Urszula Chajewska.
\newblock Neural graphical models, 2023.
\newblock URL \url{https://arxiv.org/abs/2210.00453}.

\bibitem[Stekhoven and B{\"u}hlmann(2012)]{stekhoven2012missforest}
Daniel~J. Stekhoven and Peter B{\"u}hlmann.
\newblock Miss{F}orest—non-parametric missing value imputation for mixed-type data.
\newblock \emph{Bioinformatics}, 28\penalty0 (1):\penalty0 112--118, 2012.

\bibitem[Suen and Chen(2023)]{suen2023modeling}
Daniel Suen and Yen-Chi Chen.
\newblock Modeling missing at random neuropsychological test scores using a mixture of binomial product experts.
\newblock 2023.
\newblock URL \url{https://arxiv.org/abs/2310.09384}.

\bibitem[Tchetgen et~al.(2018)Tchetgen, Wang, and Sun]{tchetgen2018discrete}
Eric J.~Tchetgen Tchetgen, Linbo Wang, and BaoLuo Sun.
\newblock Discrete choice models for nonmonotone nonignorable missing data: Identification and inference.
\newblock \emph{Statistica Sinica}, 28\penalty0 (4):\penalty0 2069, 2018.

\bibitem[Tian(2015)]{tian2015missing}
Jin Tian.
\newblock Missing at random in graphical models.
\newblock In \emph{Artificial Intelligence and Statistics}, pages 977--985. PMLR, 2015.

\bibitem[{Van Buuren}(2018)]{vanbuuren}
Stef {Van Buuren}.
\newblock \emph{Flexible Imputation of Missing Data}.
\newblock CRC Press, 2018.

\bibitem[van Buuren and Groothuis-Oudshoorn(2011)]{JSSv045i03}
Stef van Buuren and Karin Groothuis-Oudshoorn.
\newblock mice: Multivariate imputation by chained equations in {R}.
\newblock \emph{Journal of Statistical Software}, 45\penalty0 (3):\penalty0 1–67, 2011.

\bibitem[Van~der Vaart(2000)]{van2000asymptotic}
Aad~W Van~der Vaart.
\newblock \emph{Asymptotic statistics}, volume~3.
\newblock Cambridge university press, 2000.

\bibitem[Vansteelandt et~al.(2006)Vansteelandt, Goetghebeur, Kenward, and Molenberghs]{vansteelandt2006ignorance}
Stijn Vansteelandt, Els Goetghebeur, Michael~G. Kenward, and Geert Molenberghs.
\newblock Ignorance and uncertainty regions as inferential tools in a sensitivity analysis.
\newblock \emph{Statistica Sinica}, pages 953--979, 2006.

\bibitem[Weintraub et~al.(2018)Weintraub, Besser, Dodge, Teylan, Ferris, Goldstein, Giordani, Kramer, Loewenstein, Marson, et~al.]{weintraub2018version}
Sandra Weintraub, Lilah Besser, Hiroko~H. Dodge, Merilee Teylan, Steven Ferris, Felicia~C. Goldstein, Bruno Giordani, Joel Kramer, David Loewenstein, Dan Marson, et~al.
\newblock Version 3 of the alzheimer disease centers’ neuropsychological test battery in the uniform data set ({UDS}).
\newblock \emph{Alzheimer Disease \& Associated Disorders}, 32\penalty0 (1):\penalty0 10--17, 2018.

\bibitem[Yoon et~al.(2018)Yoon, Jordon, and Schaar]{yoon2018gain}
Jinsung Yoon, James Jordon, and Mihaela Schaar.
\newblock Gain: Missing data imputation using generative adversarial nets.
\newblock In \emph{International Conference on Machine Learning}, pages 5689--5698. PMLR, 2018.

\bibitem[You et~al.(2020)You, Hu, Wang, Li, Liang, and Ding]{you2020handling}
Jian You, Sheng Hu, Yu~Wang, Jiliang Li, Jun Liang, and Zhen Ding.
\newblock Handling missing data with graph representation learning.
\newblock In \emph{Proceedings of the 2020 International Conference on Learning Representations}, 2020.

\bibitem[Zhou et~al.(2010)Zhou, Little, and Kalbfleisch]{zhou2010block}
Yan Zhou, Roderick~J.A. Little, and John~D. Kalbfleisch.
\newblock Block-conditional missing at random models for missing data.
\newblock \emph{Statistical Science}, 25\penalty0 (4):\penalty0 517--532, 2010.

\end{thebibliography}

\end{document}